\documentclass{ws-mplb}

\begin{document}

\markboth{A.~M.~Dikand\'e et al.}{Dynamics of bisolitonic matter waves under gravity}

%
\catchline{}{}{}{}{}
%

\title{Dynamics of bisolitonic matter waves in a Bose-Einstein condensate subjected to an atomic beam splitter and gravity}

\author{Alain~Mo\"ise~Dikand\'e\footnote{Author for correspondence: adikande@ictp.it},~Isaiah~Ndifon~Ngek\footnote{wanndifon@mycin.net}~and Joseph~Ebobenow}

\address{Laboratory of Research on Advanced Materials and Nonlinear Sciences (LaRAMaNS), Department of Physics, Faculty of Science, University of Buea P.O. Box 63 Buea, Cameroon.}

\maketitle

\begin{history}
\received{(Day Month Year)}
\revised{(Day Month Year)}
\end{history}

\begin{abstract}
A theoretical scheme for an experimental implementation involving bisolitonic matter waves from an attractive Bose-Einstein condensate, is considered within the framework of a non-perturbative approach to the associate Gross-Pitaevskii equation. The model consists of a single condensate subjected to an expulsive harmonic potential creating a double-condensate structure, and a gravitational potential that induces atomic exchanges between the two overlapping post condensates. Using a non-isospectral scattering transform method, exact expressions for the bright-matter-wave bisolitons are found in terms of double-lump envelopes with the co-propagating pulses displaying more or less pronounced differences in their widths and tails depending on the mass of atoms composing the condensate.
\end{abstract}


\section{Introduction}
Outstanding progress has been made in the domain of laser physics thanks to the discovery of Bose-Einstein condensation~\cite{anders1,davis1}, a phenomenon observed in Bose liquids upon cooling and preparation in optical confinements~\cite{pit1}. Atom lasers~\cite{giov,pezze,pezze1} thus consist of coherent matter-wave structures that recently turned to be ideal inputs in quantum interferometry metrology such as the double-slit~\cite{shin,schumm} and Michelson-Morley~\cite{wang} interferometries. Given the robustness of their solitonic shapes, matter-wave lasers also represent an enormous potential for holographic applications involving coherent beams of optical fields which are non-destructive under interferences, wave mixings and many-body recombination processes~\cite{giov,pezze,pezze1}. \\
The present work addresses a particularly ineresting matter-wave laser implementation in which the input atom laser is a double-lump pulse, obtained by splitting a single Bose-Einstein condensate(BEC) into two parts. Such implementation has relevant application in the double-slit holography~\cite{pezze,pezze1} where two overlapping condensates are created by splitting one single condensate, e.g. by a far-off-resonance laser barrier~\cite{andrews}, leading to two atomic populations in nearly degenerate hyperfine levels~\cite{stamper1}. In many past theoretical studies devoted to the last subject, the far-off-resonance splitting is described by a double-well potential~\cite{shenoy}. A main motivation for the choice of a double-well potential is the presence of a potential hump that cuts the condensate into two slices, and of two minima allowing harmonic confinement of each of the two overlapping condensates. However, in a strict sense the mean-field picture predicts that the governing mechanism of formation of matter-wave solitons is the interparticle interaction and the trap interefers as a post process, the role of which is to confine matter-wave solitons thus permitting full control and experimental manipulations of the resulting laser beam. Therefore, in studying the far-off-resonance splitting process of a single BEC it is enough considering only the hump part of the double-well potential, a part which can well be represented by an harmonic expulsive (i.e. anti-trap) potential~\cite{kengne,dika1,liua,zhang}. \\
Matter-wave lasers are relatively more sensitive to a gravitational field than the classic photon laser for the matter-wave laser source is an excited population of atoms. Particularly in the case of condensates consisting of vapor atoms the free fall stands for a relevant process promoting atomic exchanges between two overlapping condensates, and this process is expected to be dominant over scattering processes such as thermal feedings and many-body recombinations in the case of heavy atoms BECs. From the theoretical standpoint, the influence of gravity on Bose systems has been discussed, firstly in the context of non-interacting particles~\cite{landau,wadati1} and more recently in the case of a system of weakly interacting Bose particles~\cite{wadati2}. In this last work, following the WKB approximation the authors obtained that the condensate wavefunction could be approximated by solutions of a perturbed linear Schr\"odinger equation. On the basis of this assumption, they derived expressions of the condensate wavefunction as combinations of Bessel functions. \\
In this work we shall develop a non-perturbative theory for the mean-field dynamics of an atomic population consisting of two condensates held together by an harmonic anti-trap potential and exchanging atoms by means of the gravitational field. As already indicated, a BEC phenomenon of this kind has established physical importance~\cite{mewes,bloch,coq,hope,mori,shimizu} and it will emerge below that in the specific context of BEC systems with relatively heavy atoms, the gravity can strongly affect both shape profiles and the dynamics of pulses in the matter-wave bisoliton created by the anti-trap~\cite{coq,hope} so that perturbative techniques, including variational methods, are very limited in their description. 
\section{Model and analytical bright-matter bisoliton solutions}
Consider the following Gross-Pitaevskii equation, which governs the groundstate wavefunction of a BEC system in an anti-trap potential and subjected to the gravitational field:
\begin{equation}
i\hbar\,\psi_t= -\frac{\hbar^2}{2m}\,\nabla^2 \psi + \left[ P_a\,z - m\,\omega_{\parallel}^2\,z^2 - 2\xi \mid \psi \mid^2\right] \psi,      \label{a1}
\end{equation}
where $\xi$ is the (positive) mean-field interatomic interaction coefficient related to the $s-wave$ scattering length, $m$ is the atomic mass and $P_a=mg$ the associate weight with $g$ the acceleration of gravity. $\omega_{\parallel}$ is the characteristic frequency for atomic oscillations along the axis of graviational force, as we are interested in the dynamics of the above system along the last axis~\cite{hope} we shall assume $\nabla^2\equiv \partial^2/\partial z^2$. \\
To gain the best formulation of the dynamics of BEC systems described by the mean-field Eq.~(\ref{a1}), we follow the Inverse-Scattering-Transform (IST) method which offers an interesting non-perturbative framework for the study of exactly integrable nonlinear partial differential equations. To this last end, remark to start that when the gravity and the expulsive potential are absent Eq.~(\ref{a1}) reduces to the homogeneous Nonlinear Schr\"odinger (NLS) equation which, within the framework of the IST method, gives rise to an isospectral Lax-pair-type eigenvalue problem with universal conservation laws~\cite{dodd}. However, in the presence of perturbation, this equation can become a rather complicated mathematical problem and is rarely exactly integrable. Nevertheless, exploiting recent developments in the treatment of non-isospectral Lax-pair eigenvalue problems~\cite{kong1,kong2,qiao,pilar,ning1,zuo}, non-universal conservation laws can be derived for a certain class of perturbations such that the associate NLS equations are exactly integrable. In the specific case of Eq.~(\ref{a1}) such conservation laws have been constructed in the cases $P_a\neq 0$ but $\omega_{\parallel}=0$~\cite{liu1}, $P_a=0$ but $\omega_{\parallel}\neq 0$~\cite{bala}, $P_a\neq 0$ and $\omega_{\parallel}\neq 0$~\cite{dika1}. Following the method for this last case, the two pairs of linear eigenvalue problems associate with the perturbed NLS Eq.~(\ref{a1}) are defined as:
\begin{eqnarray}
u_z + i\,\lambda\,u&=& \psi(z, t)\,v, \nonumber \\ 
v_z - i\,\lambda\,v&=& -\psi^{\star}(z, t)\,u, \label{a2}
\end{eqnarray}
\begin{eqnarray}
u_t - A\,u - B\,v&=&0, \nonumber \\ 
v_t - C\,u + A\,v&=&0,  \label{a3}
\end{eqnarray}
where $A$, $B$, $C$ are coefficients of the matrix associate with the second linear system~(\ref{a3})~\cite{liu1,bala}, $u(z, t)$ and $v(z, t)$ are space-time dependent functions forming elements of a two-component wavevector while $\lambda$ is the associate eigenvalue. The quantity $\psi(x, t)$ is the solution wanted, but now plays the role of IST's scattering potential. According to the general rule~\cite{dodd,liu1,bala} of the IST technique, the appropriate space-time evolutions of the matrix elements $A$, $B$ and $C$ should be determined unambiguously provided conservation laws for the set~(\ref{a2}) and (\ref{a3}) are explicitely formulated. These conservation laws follow from the condition of compatibility between~(\ref{a2}) and (\ref{a3}) corresponding to a derivation of~(\ref{a2}) with respect to $t$ and~~(\ref{a3}) with respect to $z$ and comparing the four resulting equations. This leads to the following set of three coupled evolution equations:   
\begin{eqnarray}
A_z + i\,\lambda_t&=&-C\,\psi(z, t) - B\,\psi^{\star}(z, t),  \label{aa1}, \\
(\ln B)_z + 2i\,\lambda &=&\psi_t(z, t)/B - 2A\,\psi(z, t)/B, \label{aa2} \\
(\ln C)_z - 2i\,\lambda &=&-\psi^{\star}_t(z, t)/C - 2A\,\psi^{\star}(z, t)/C, \label{aa3} 
\end{eqnarray}
where the eigenvalue $\lambda$ is assumed space-time dependent as a result of the perturbation in the NLS~(\ref{a1}), hence the non-isospectral character of the IST eigenvalue problem. \\
As we are interested in the matter-wave structures with pulse shape, we need to operate a judicious choice of the scattering potential. For the shape profile of current interest the scattering potential must vanish asymptotically as $z\rightarrow \pm \infty$ at any propagation time $t$, meaning that a reflectionless potential stands for an appropriate choice. In the specific case of the perturbed NLS Eq.~(\ref{a1}), a best representation for the initial soliton solution with pulse shape is:  
\begin{equation}
\psi(z, t=0)= -2\lambda(z, t=0)\,sech\left[2 \int_{z_0}^z{\lambda(z', t=0)dz'}\right], \label{a3b}
\end{equation}
where $\lambda(z, t=0)$, the eigenvalue of the IST problem at the initial time $t=0$, is assumed to vary in both space and time in accordance with the non-isospectral character of the associate eigenvalue problem. To put the last remark on a firm ground, it suffices to notice that by combining the compatibility relations~(\ref{aa1})-(\ref{aa3}) with the generic equation~(\ref{a1}), we can derive a single compatibility equation in terms of the sole eigenvalue $\lambda$ i.e.:
\begin{equation}
V_z - \lambda_t  - 2(i\lambda)^2_z=0, \hskip 0.4truecm V(z)= P_a\,z - m\,\omega_{\parallel}^2\,z^2. \label{a3c}
\end{equation}
This compatibility equation indeed gives clear evidence of a dependence of $\lambda$ on the space and time coordinates, establishing thus the non-universal character of spectral parameters of the eigenvalue problems~(\ref{a2})-(\ref{a3}). It turns out that $\lambda$ can be written explicitely as $\lambda(z, t)$, and in general any solution provided by~(\ref{a3c}) must be generic of eigenvalue(s) of the IST scattering potential defined in~(\ref{a3b}). \\
Now grouping all the preceeding key considerations in the standard steps of the IST method~\cite{dika1,dodd,bala}, we arrive at the following exact one-soliton solution for the perturbed NLS Eq.~(\ref{a1}):
\begin{equation}
 \psi(z, t)=\psi_0(z, t)sech\left[2\Phi(z, t)\right]\,\exp[-i\varphi(z,t)], \hskip 0.4truecm \psi_0(z,t)= \vartheta_0\,G(z)F_i(t), \label{a4}
 \end{equation}
 where
 \begin{equation}
 \Phi(z, t)= - \frac{\sqrt{2 m \xi}}{\hbar} F_i(t)\int_{z_0}^z{G(y)dy} + \ln\sqrt{\frac{\mid \vartheta_0 \mid}{2F_i(t)}}
               + \frac{2P_a}{\sqrt{\xi}}\int_0^{t}{\Im{[(F^2(t')]}d t'}, \label{a5}
\end{equation}
\begin{equation}	       
 \varphi(z, t)=  \frac{2\sqrt{2 m \xi}}{\hbar} F_r(t)\int_{z_0}^z{G(y)dy}
+ \frac{4P_a}{\sqrt{\xi}}\int_0^{t}{\Re{[(F^2(t')]}dt'},  \label{a6}
\end{equation}   
\begin{equation}
 G(z)= \frac{1}{2\xi\sqrt{\xi}}\left[ P_a - \frac{2m\sqrt{2m}}{\hbar}\omega_{\parallel}^2\, z \right], \label{a7}
 \end{equation}
$F(t)=F_i(t) + i F_r(t)$ with
\begin{equation}
F_r(t)= \frac{1}{\lambda}\,\frac{(1 + \lambda^2)\tanh(\kappa t)}{1 + \lambda^2\tanh^2(\kappa t)}, \hskip 0.4truecm
F_i(t)= \frac{sech^2(\kappa t)}{1 + \lambda^2\tanh^2(\kappa t)}, \label{a8}
 \end{equation}
and
 \begin{equation}
 \kappa= -2\omega_{\parallel}\sqrt{m}/\hbar, \hskip 0.4truecm \lambda= 1/\kappa. \label{a9}
 \end{equation}
Note that the quantity $\vartheta_0$ in~(\ref{a4}) depends entirely on amplitudes of the conjugate pair of eigenfunctions (i.e. Jost functions) of the IST scattering problem~\cite{dika1}, hence it does not depends on space and time and can therefore readily be regarded as an arbitrary amplitude of the IST one-soliton solution. 
\section{Numerical analysis of the analytical bisoliton solution}
\begin{figure*}
\begin{minipage}[c]{0.32\textwidth}
\includegraphics[width=4.0cm]{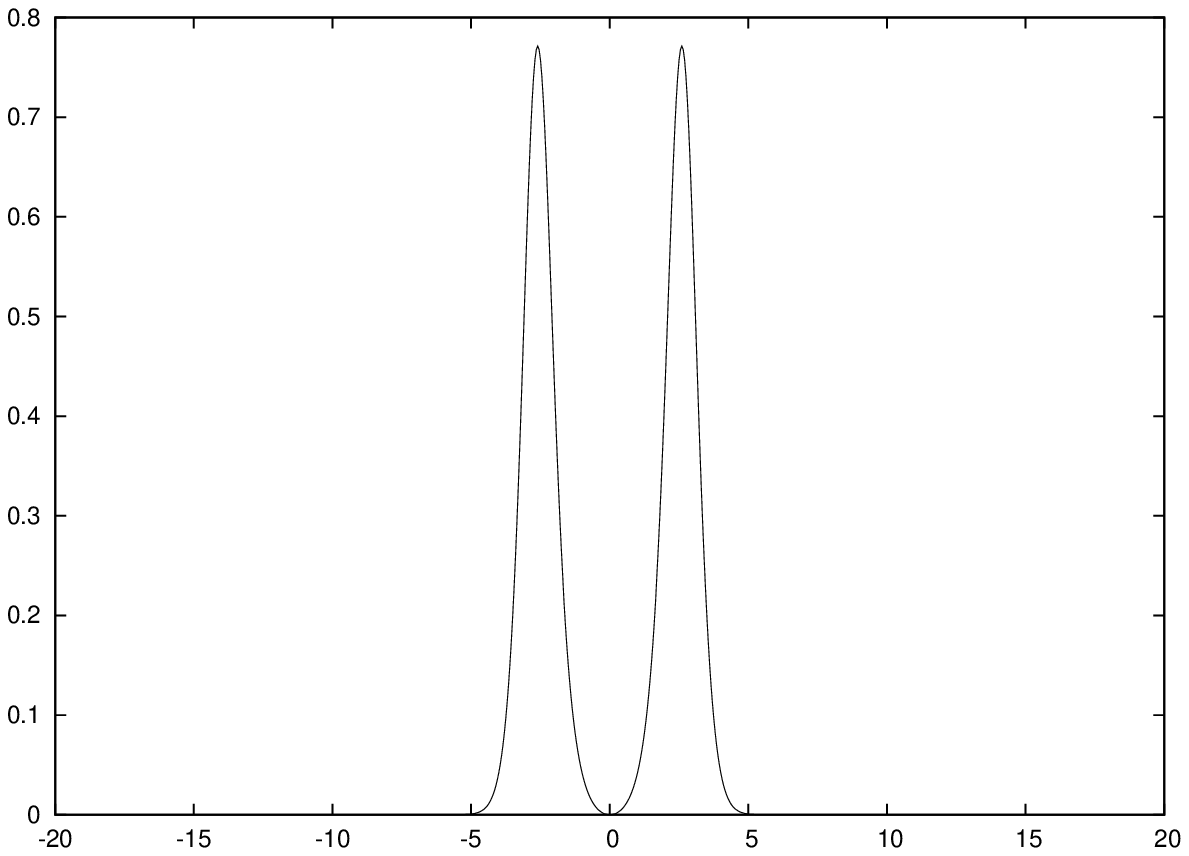}
\end{minipage}%
\begin{minipage}[c]{0.32\textwidth}
\includegraphics[width=4.0cm]{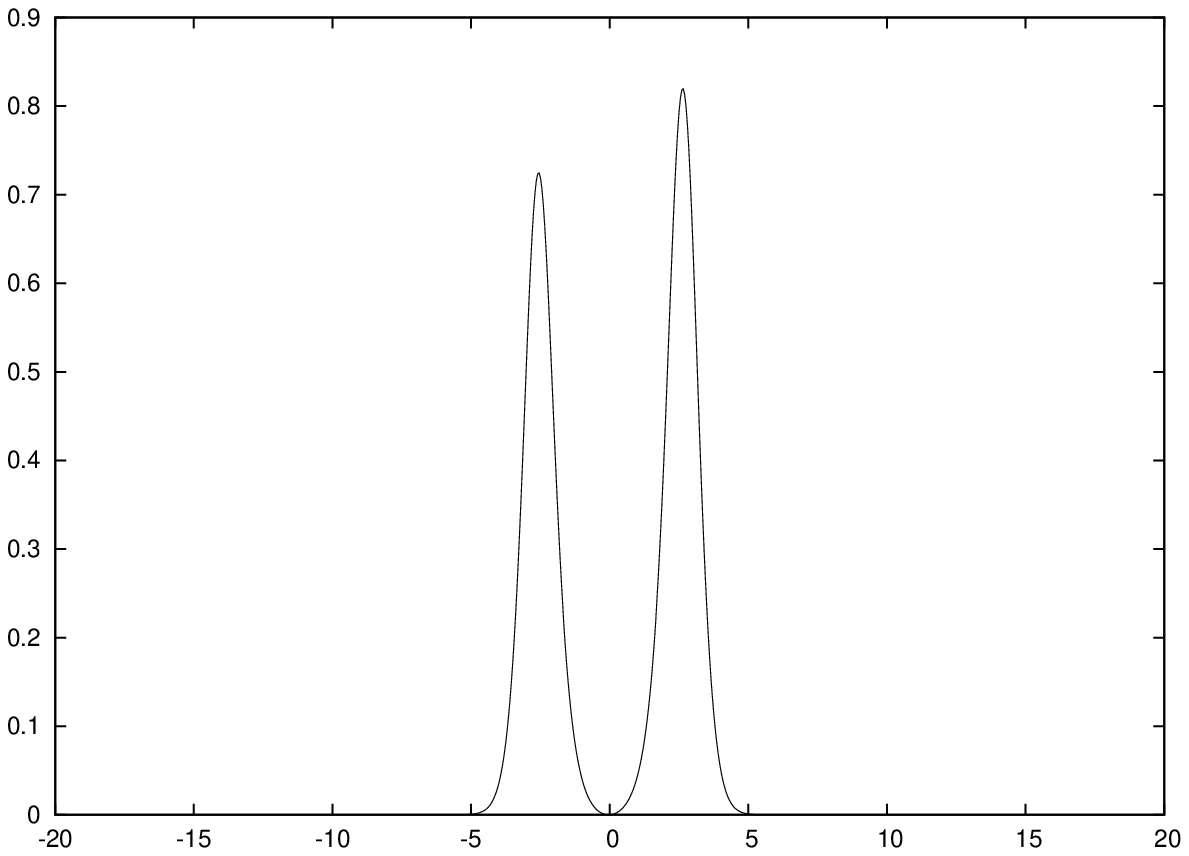}
\end{minipage}%
\begin{minipage}[c]{0.32\textwidth}
\includegraphics[width=4.0cm]{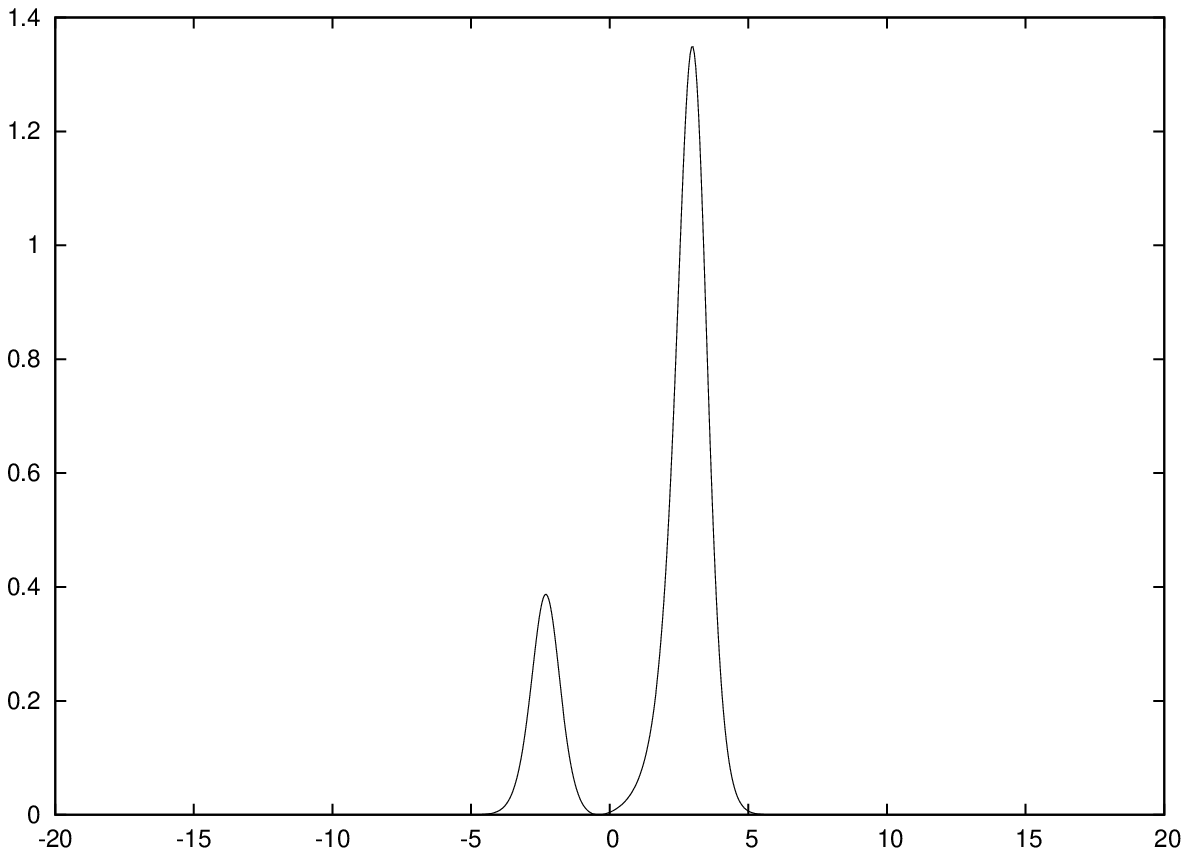}
\end{minipage} \vskip 0.4truecm
\begin{minipage}[c]{0.32\textwidth}
\includegraphics[width=4.0cm]{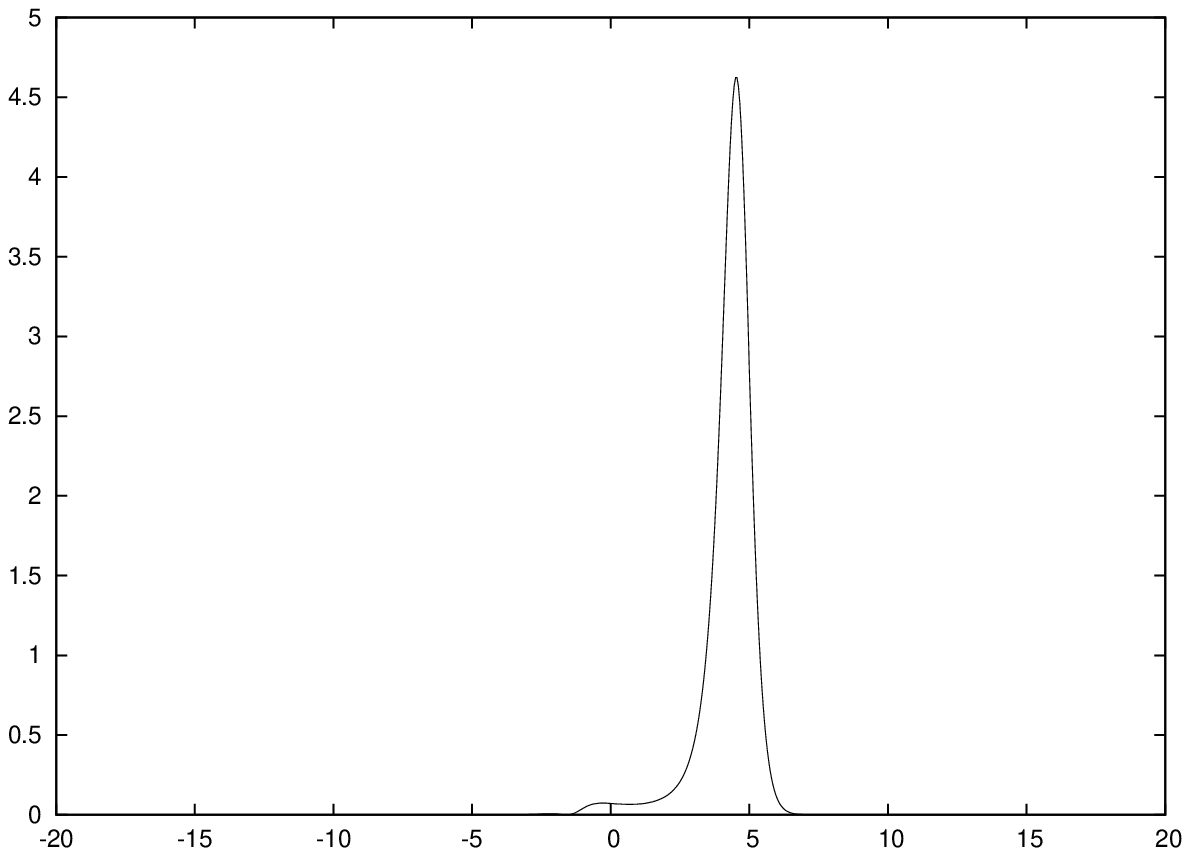}
\end{minipage} 
\begin{minipage}[c]{0.32\textwidth}
\includegraphics[width=4.0cm]{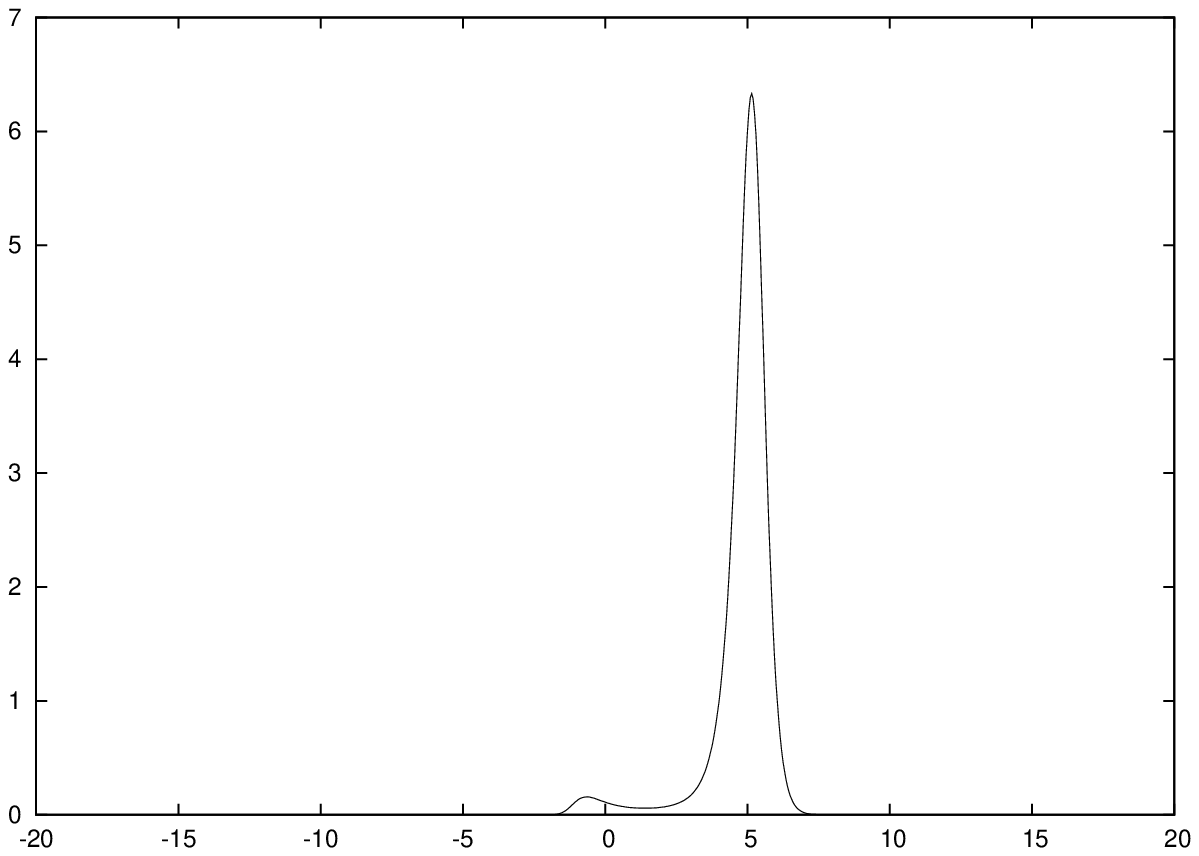}
\end{minipage}%
\begin{minipage}[c]{0.32\textwidth}
\includegraphics[width=4.0cm]{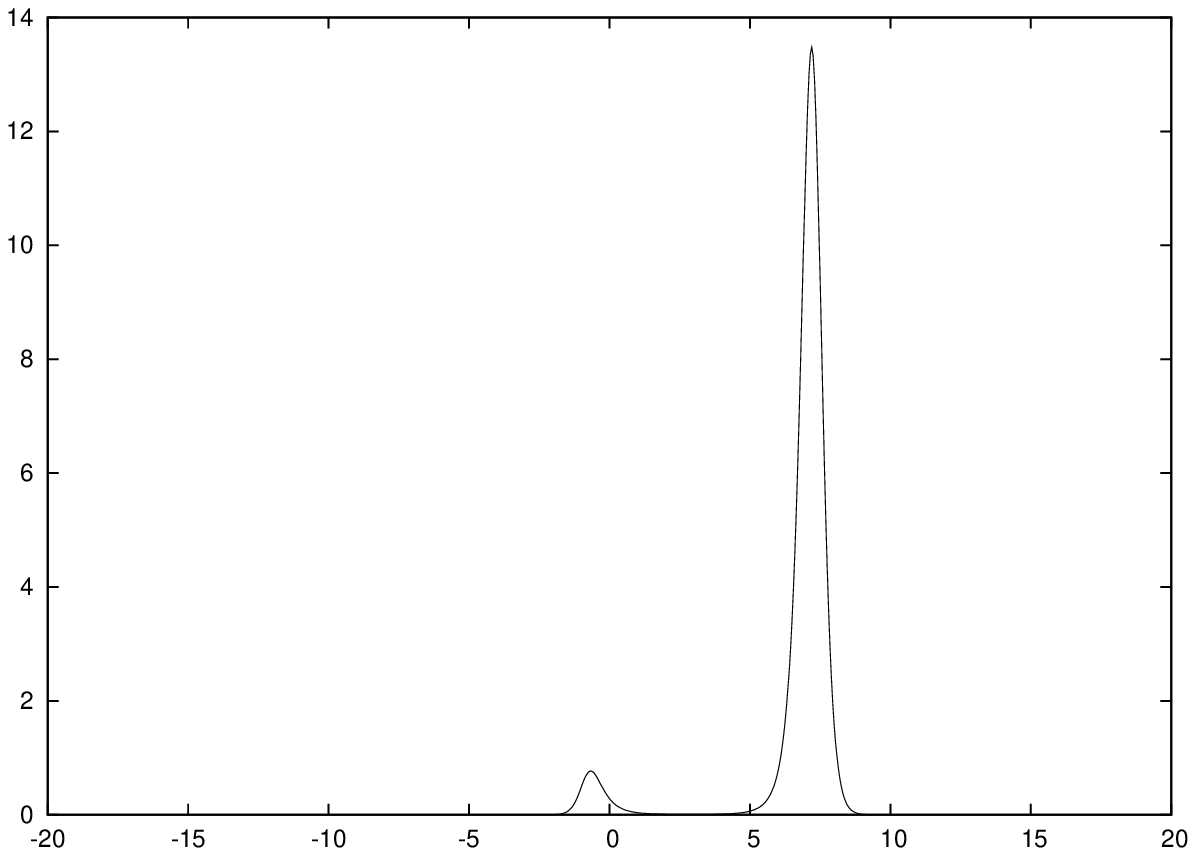}
\end{minipage} 
\caption{Amplitude $\vert \psi(z, 0)\vert$ of the initial matter-wave intensity versus position $z$. For all the six graphs $m\omega_{\parallel}^2=0.5$ and $P_a$ is varied as follows, from the top left to the last graph: $0$, $0.01$, $0.1$, $0.4$, $0.5$, $0.8$.} 
\label{figone}
\end{figure*}

\begin{figure*}
\begin{minipage}[c]{0.51\textwidth}
\includegraphics[width=5.2cm]{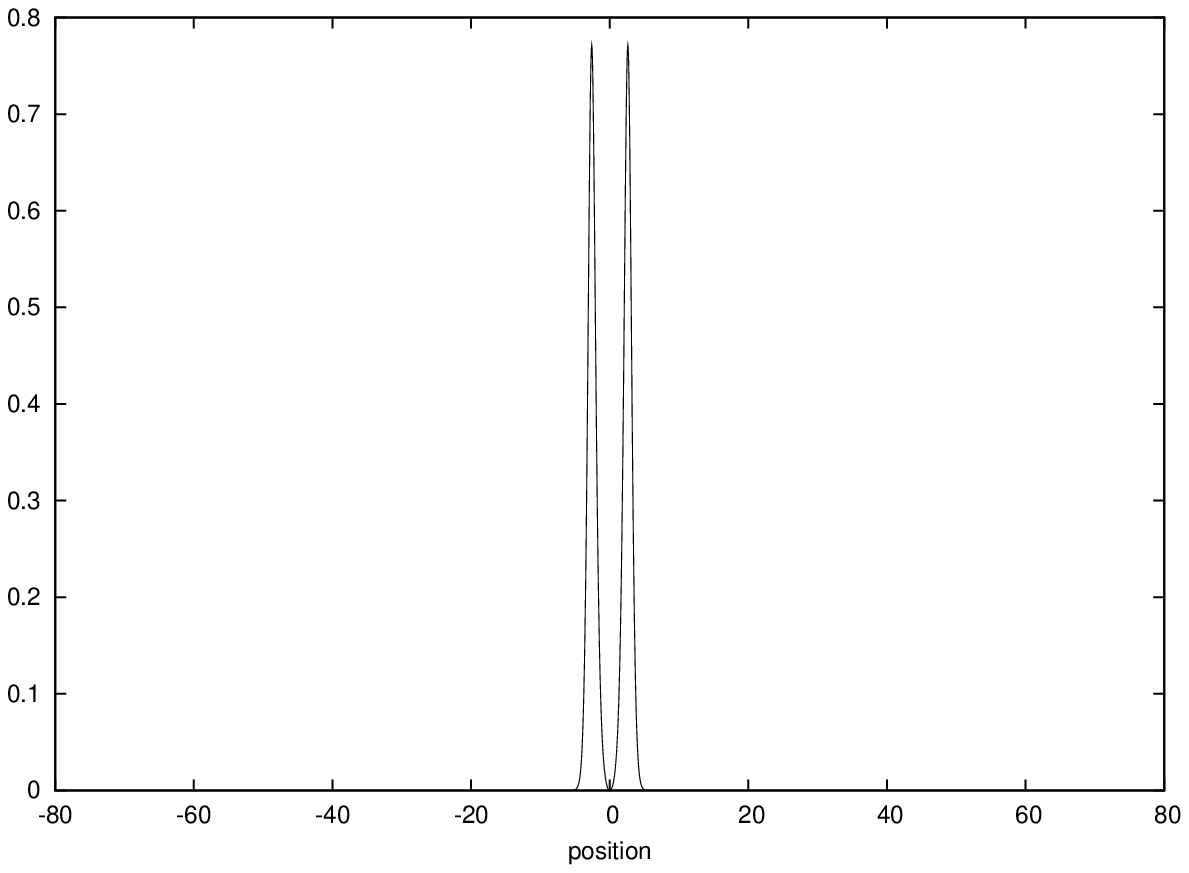}
\end{minipage}%
\begin{minipage}[c]{0.51\textwidth}
\includegraphics[width=5.2cm]{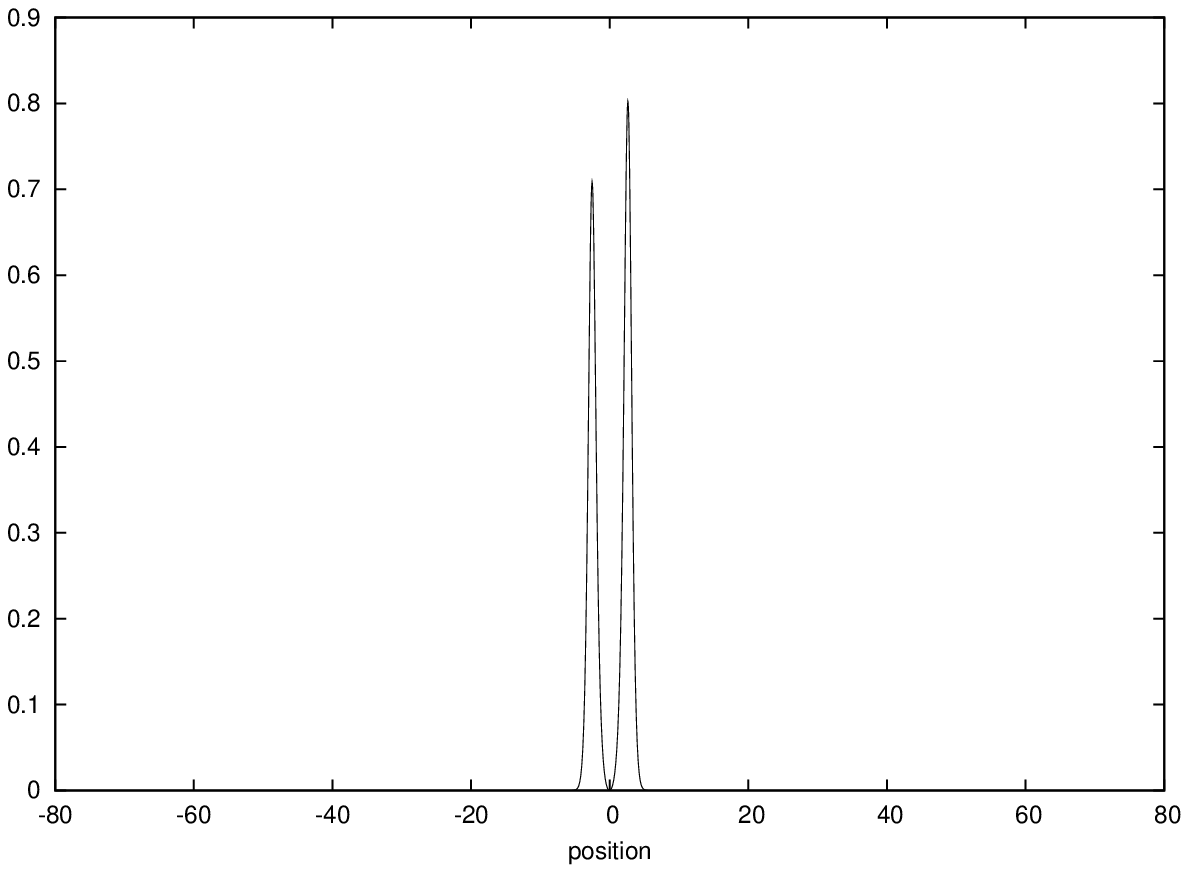}
\end{minipage}\vskip 0.3truecm 
\begin{minipage}[c]{0.51\textwidth}
\includegraphics[width=5.2cm]{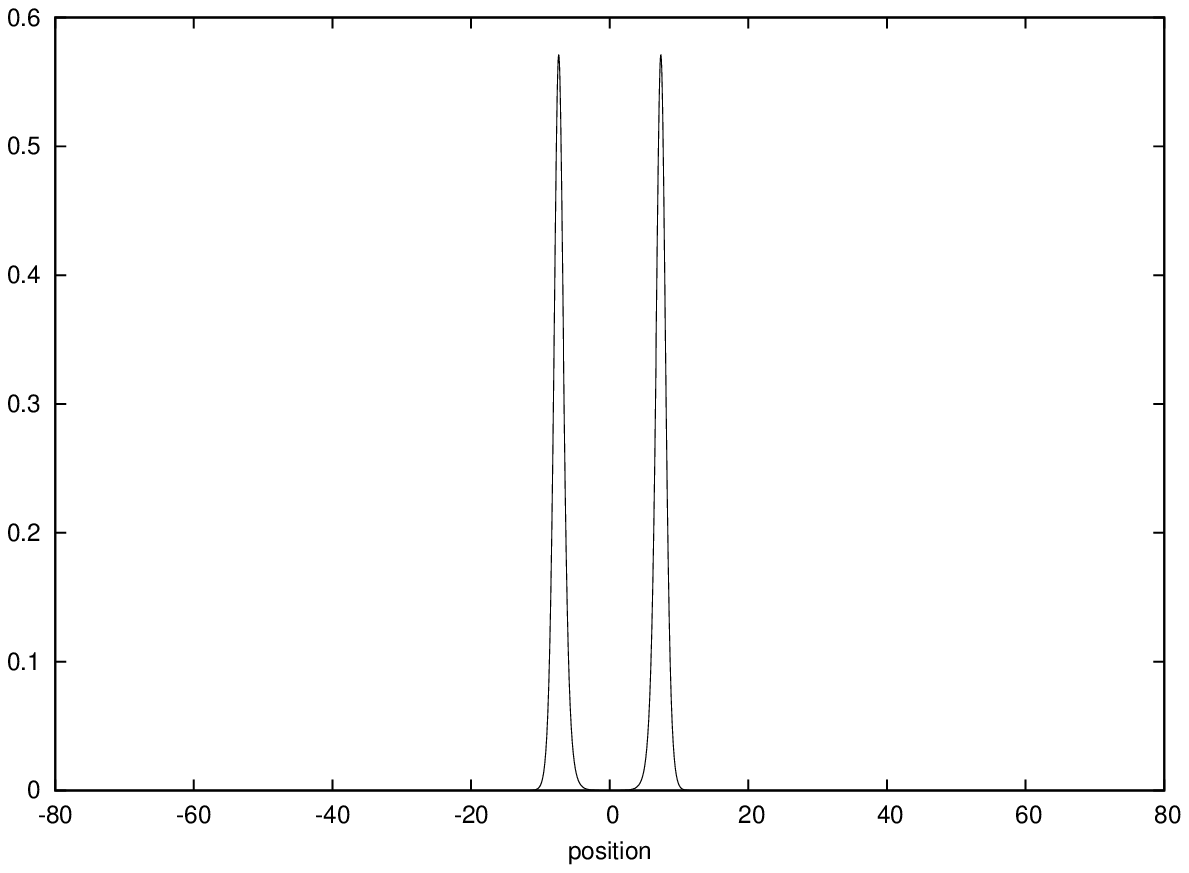}
\end{minipage}%
\begin{minipage}[c]{0.51\textwidth}
\includegraphics[width=5.2cm]{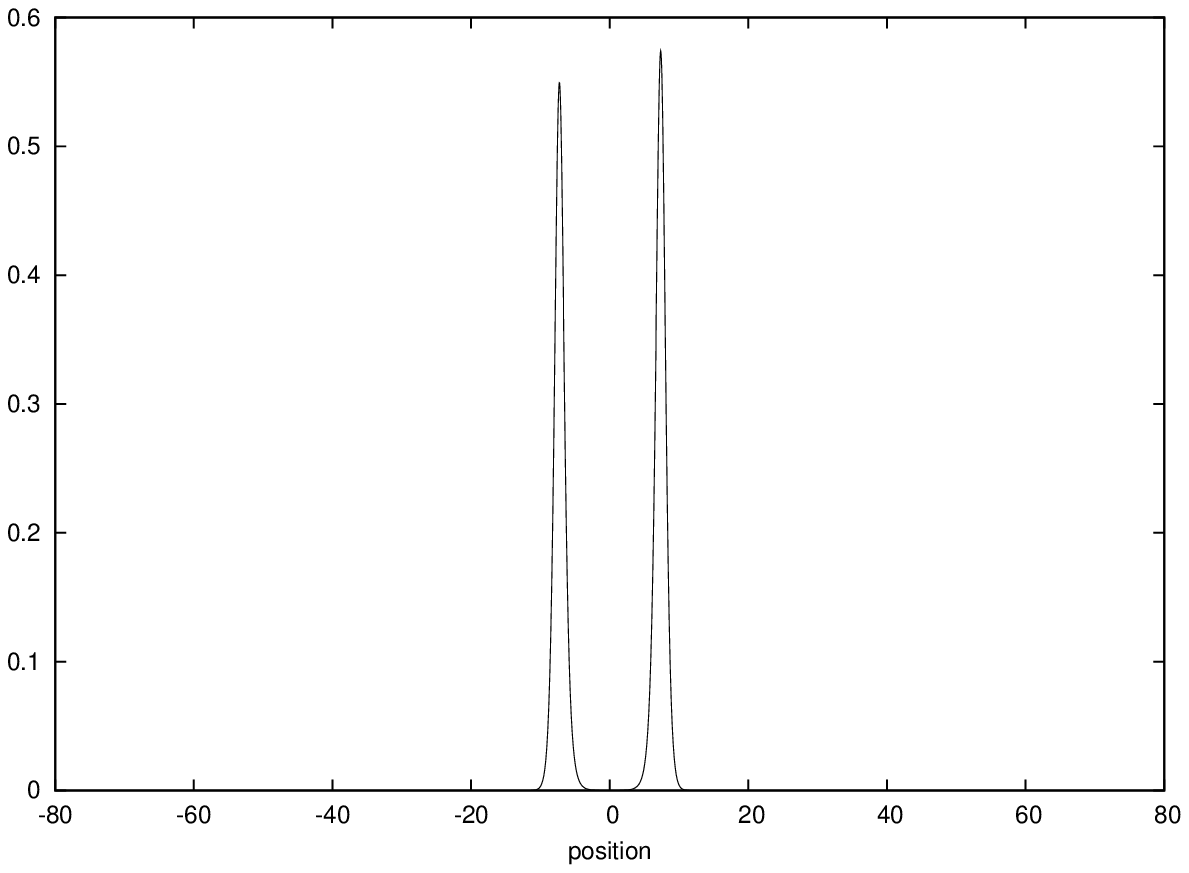}
\end{minipage}\vskip 0.3truecm 
\begin{minipage}[c]{0.51\textwidth}
\includegraphics[width=5.2cm]{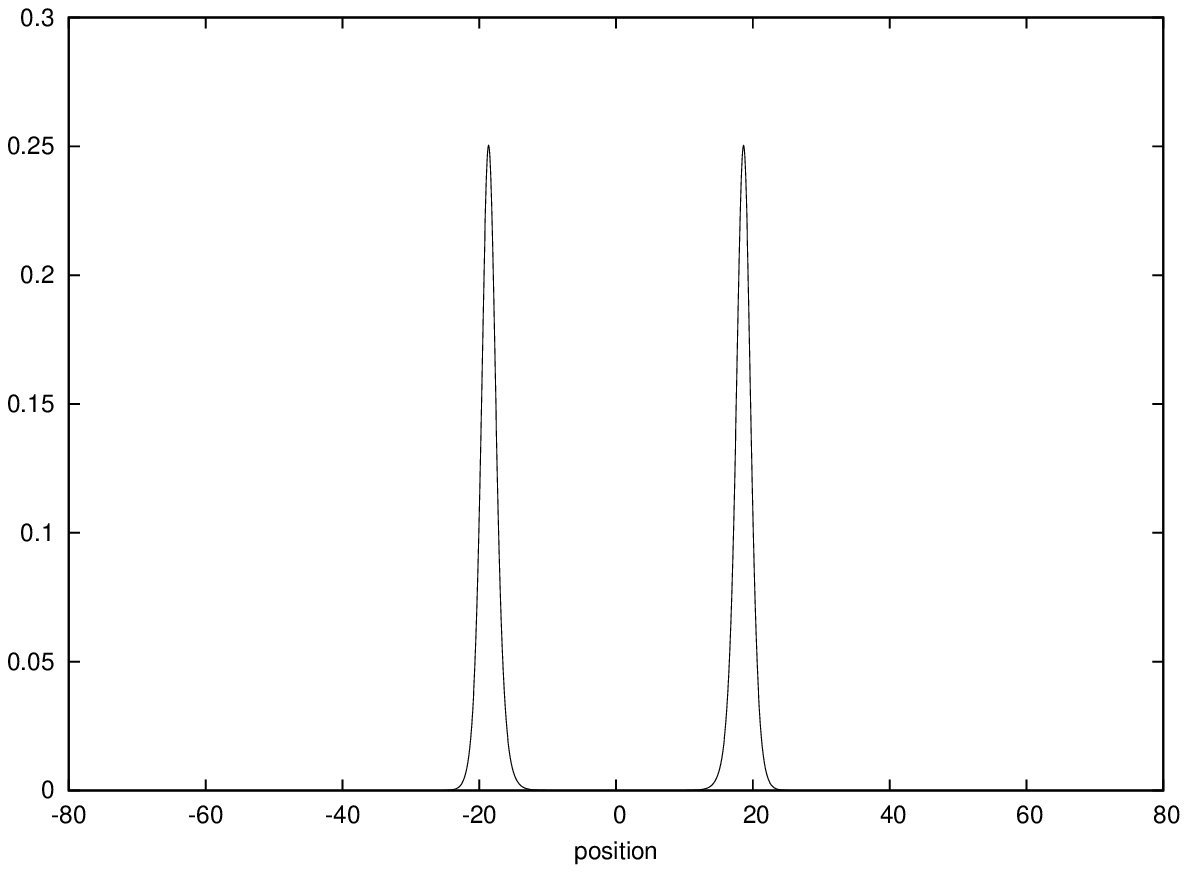}
\end{minipage}%
\begin{minipage}[c]{0.51\textwidth}
\includegraphics[width=5.2cm]{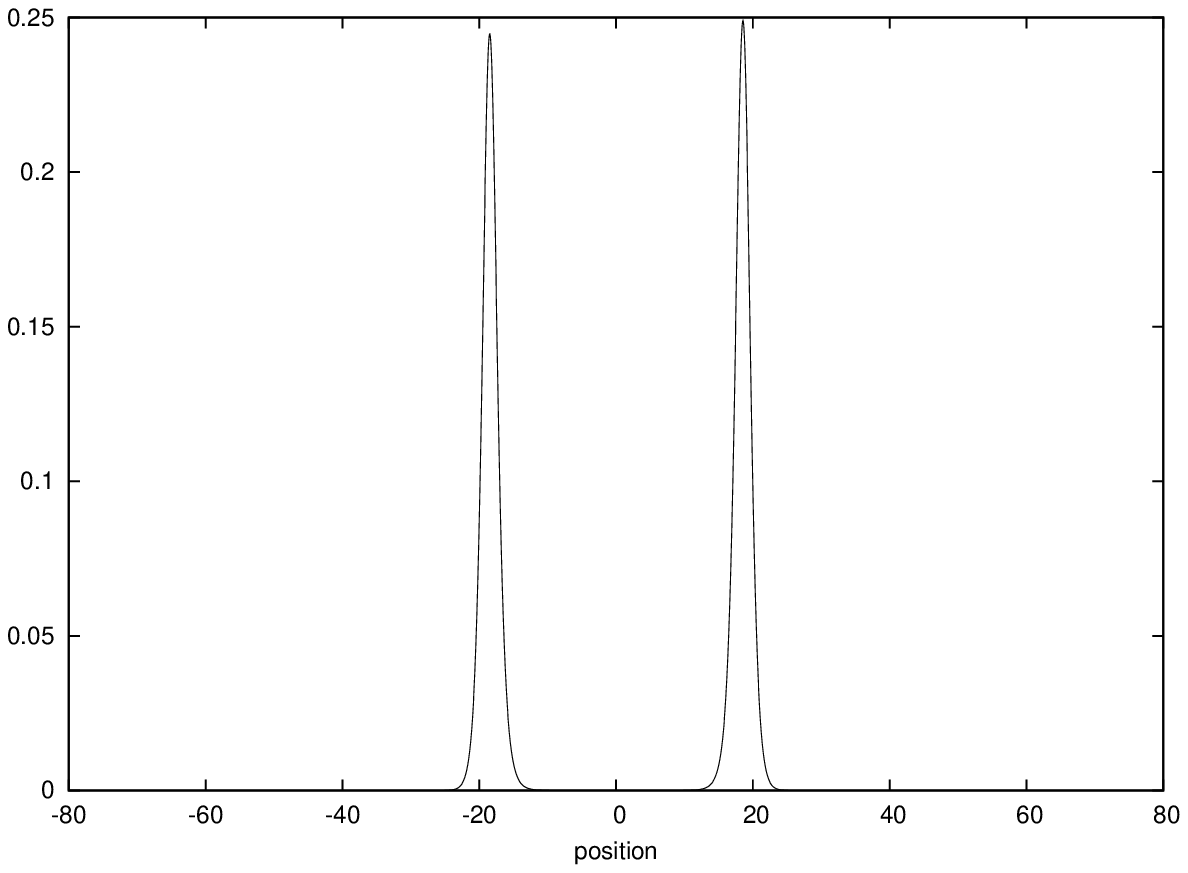}
\end{minipage}\vskip 0.3truecm 
\begin{minipage}[c]{0.51\textwidth}
\includegraphics[width=5.2cm]{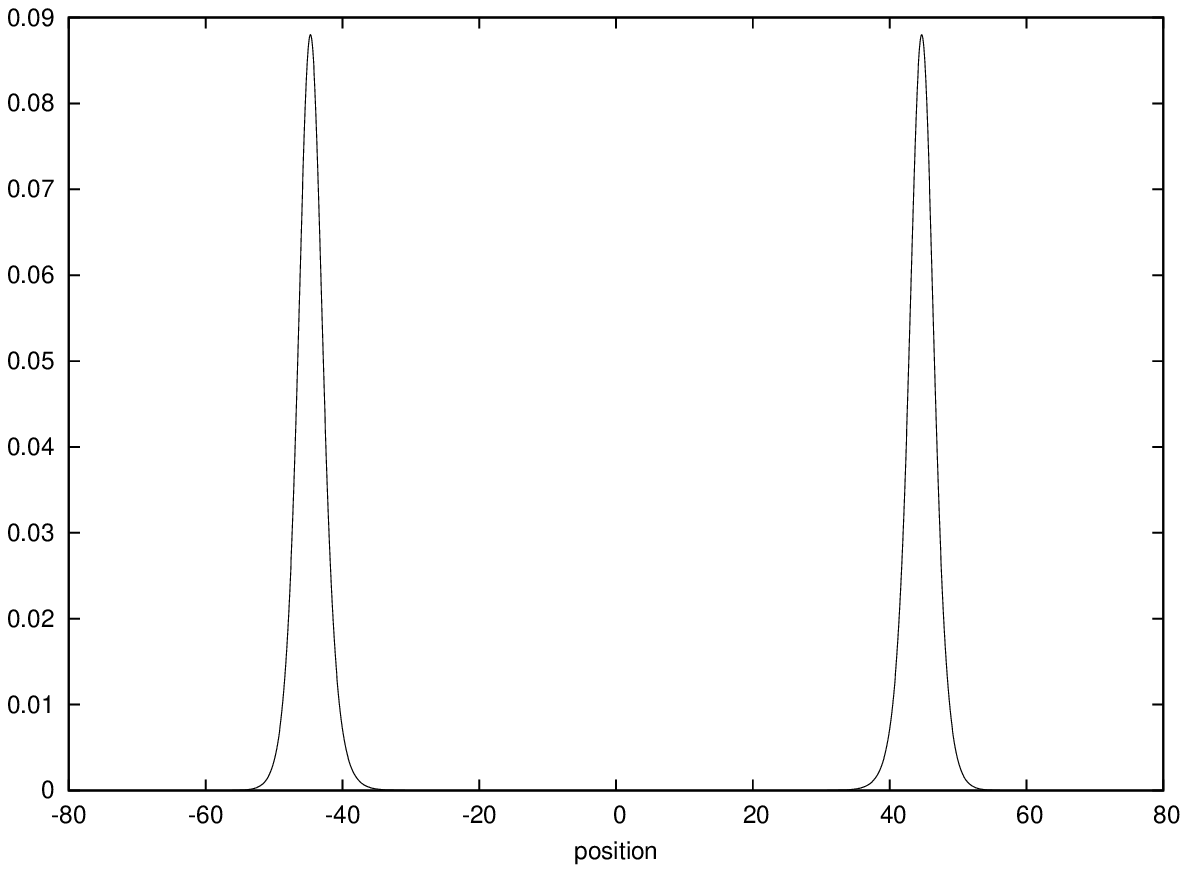}
\end{minipage}%
\begin{minipage}[c]{0.51\textwidth}
\includegraphics[width=5.2cm]{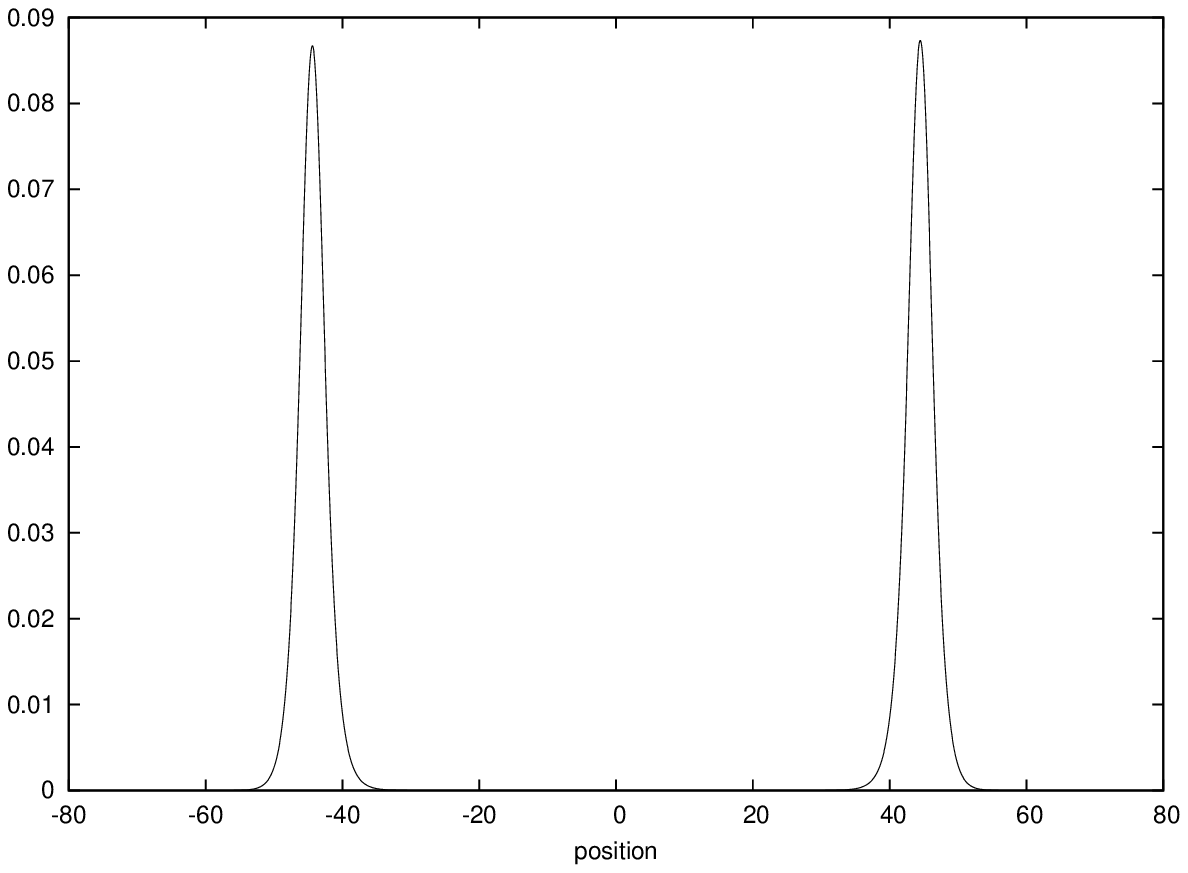}
\end{minipage} \vskip 0.3truecm
\begin{minipage}[c]{0.51\textwidth}
\includegraphics[width=5.2cm]{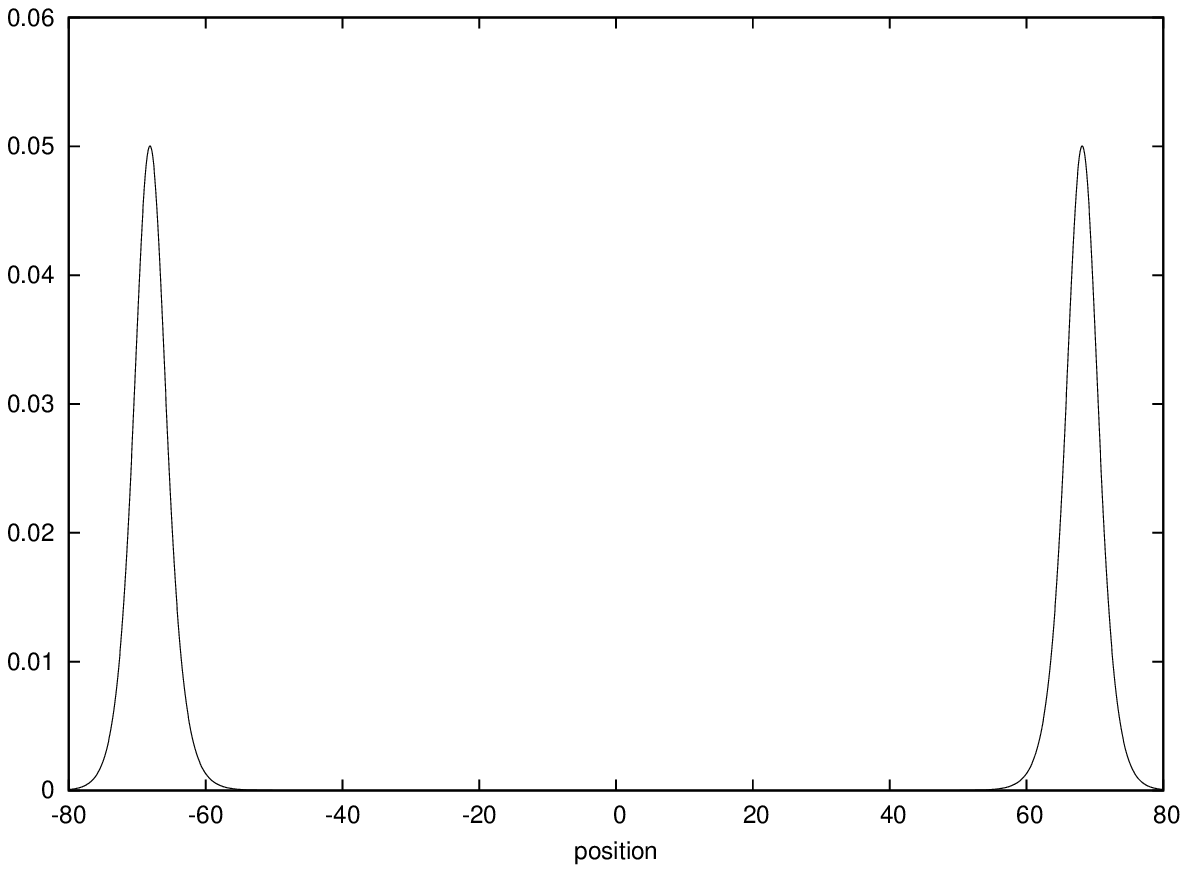}
\end{minipage}%
\begin{minipage}[c]{0.51\textwidth}
\includegraphics[width=5.2cm]{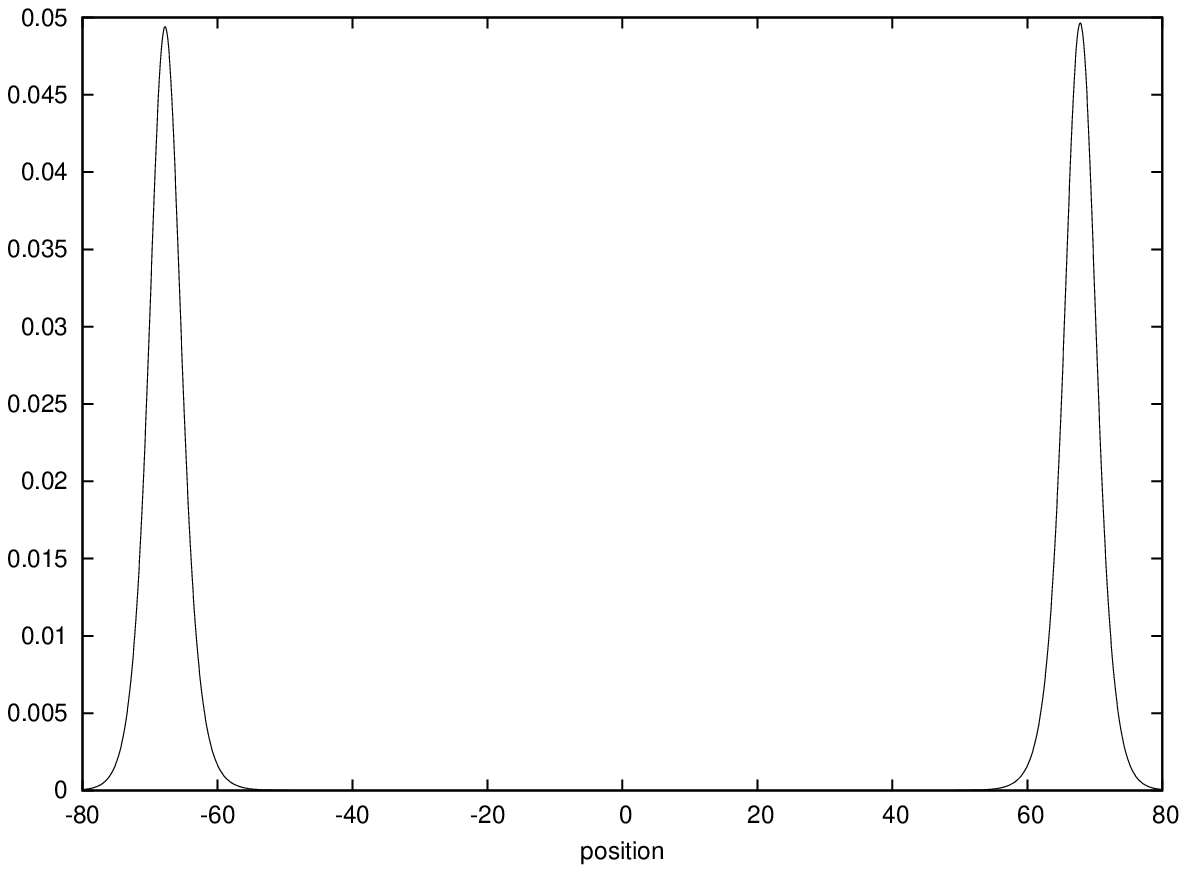}
\end{minipage}
\caption{Spatial profile of the intensity $\vert \psi(z, t)\vert$ of the matter-wave bisoliton for $P_a=0$ (left column) and $P_a=0.01$ (right column). From the top to the bottom graphs times are choosen as follows, in arbitrary units: $t=0$, $1.5$, $2.5$, $3.5$, $4$} 
\label{figtwo}
\end{figure*}

\begin{figure*}
\begin{minipage}[c]{0.51\textwidth}
\includegraphics[width=5.2cm]{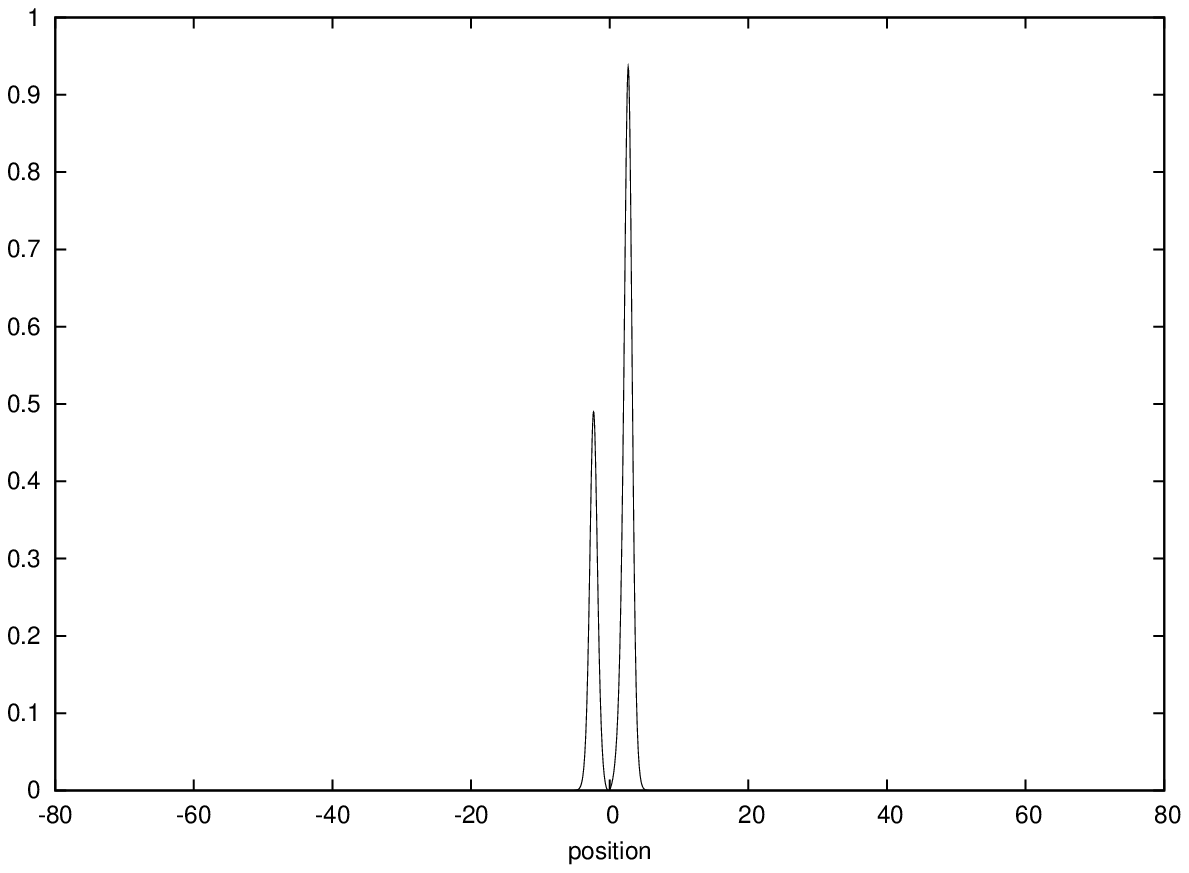}
\end{minipage}%
\begin{minipage}[c]{0.51\textwidth}
\includegraphics[width=5.2cm]{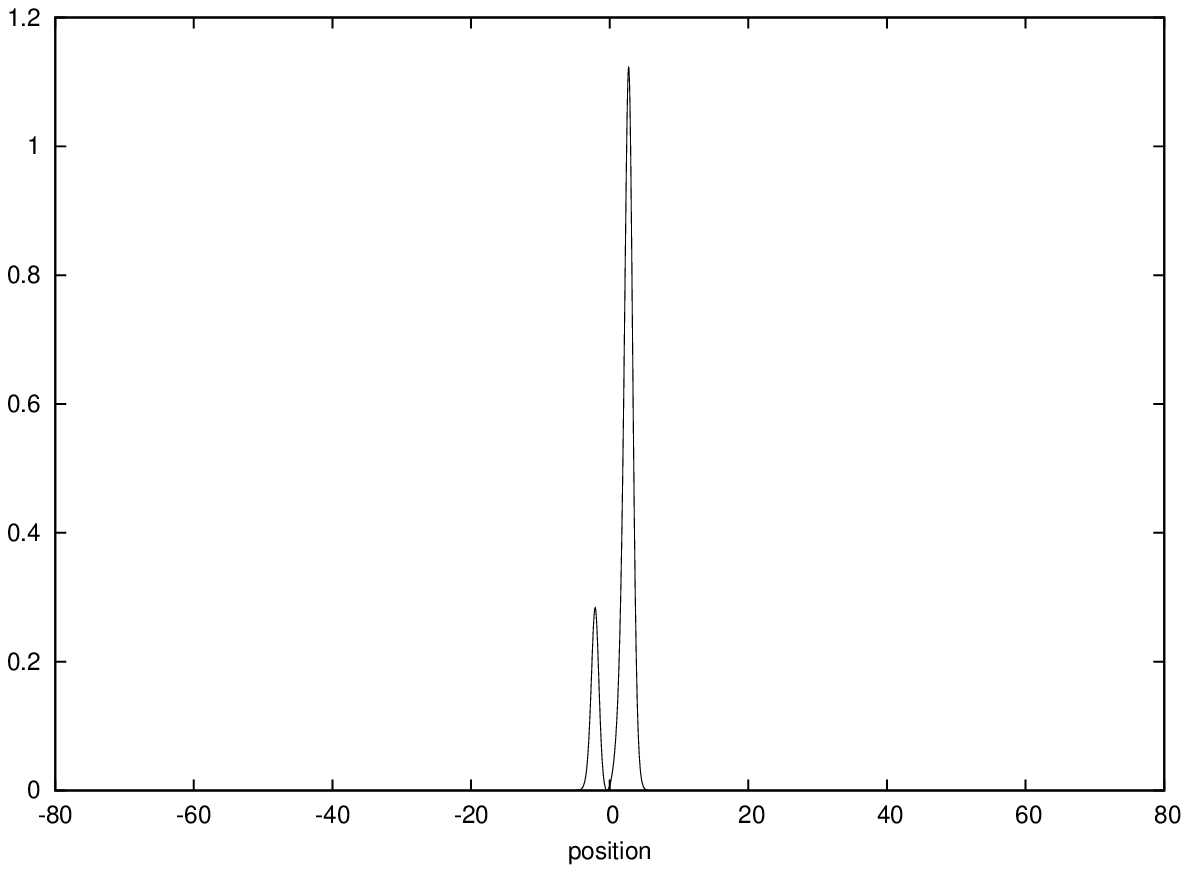}
\end{minipage}\vskip 0.4truecm
\begin{minipage}[c]{0.51\textwidth}
\includegraphics[width=5.2cm]{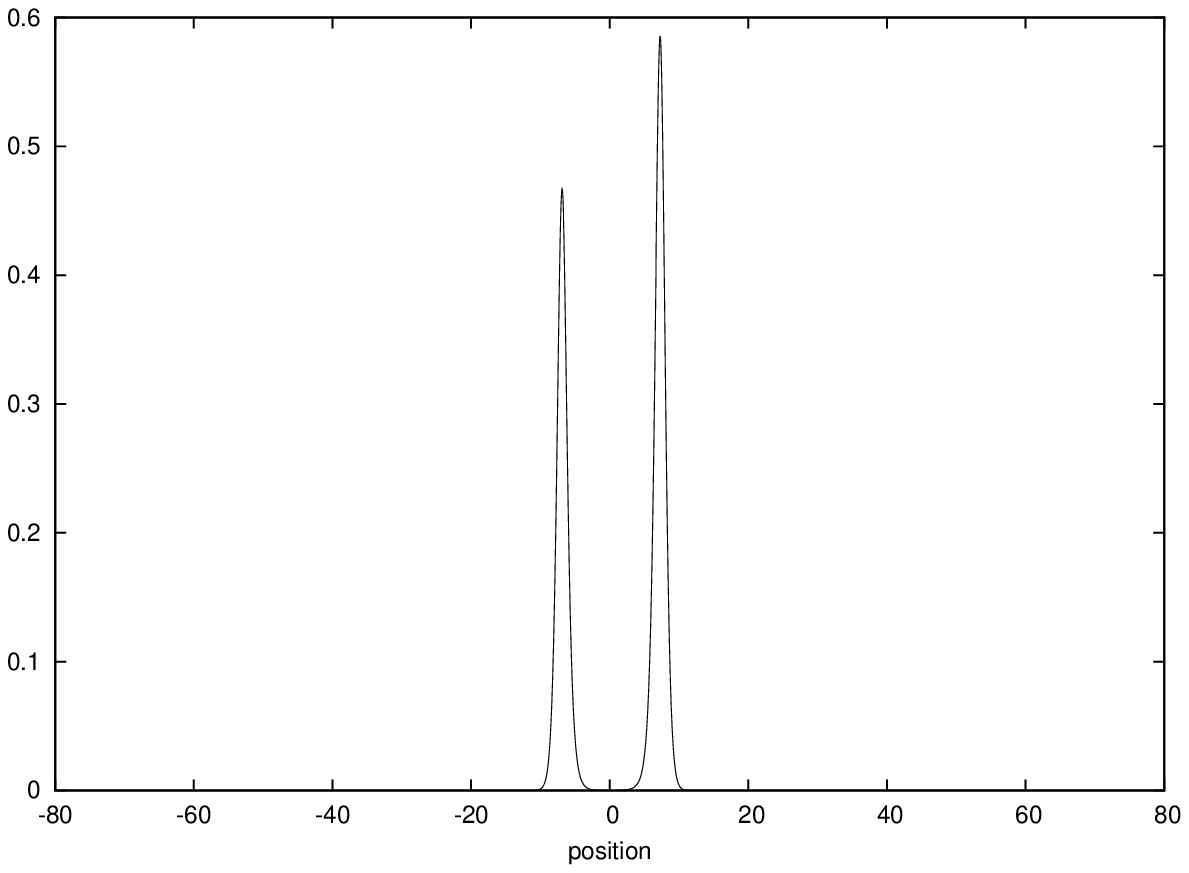}
\end{minipage}%
\begin{minipage}[c]{0.51\textwidth}
\includegraphics[width=5.2cm]{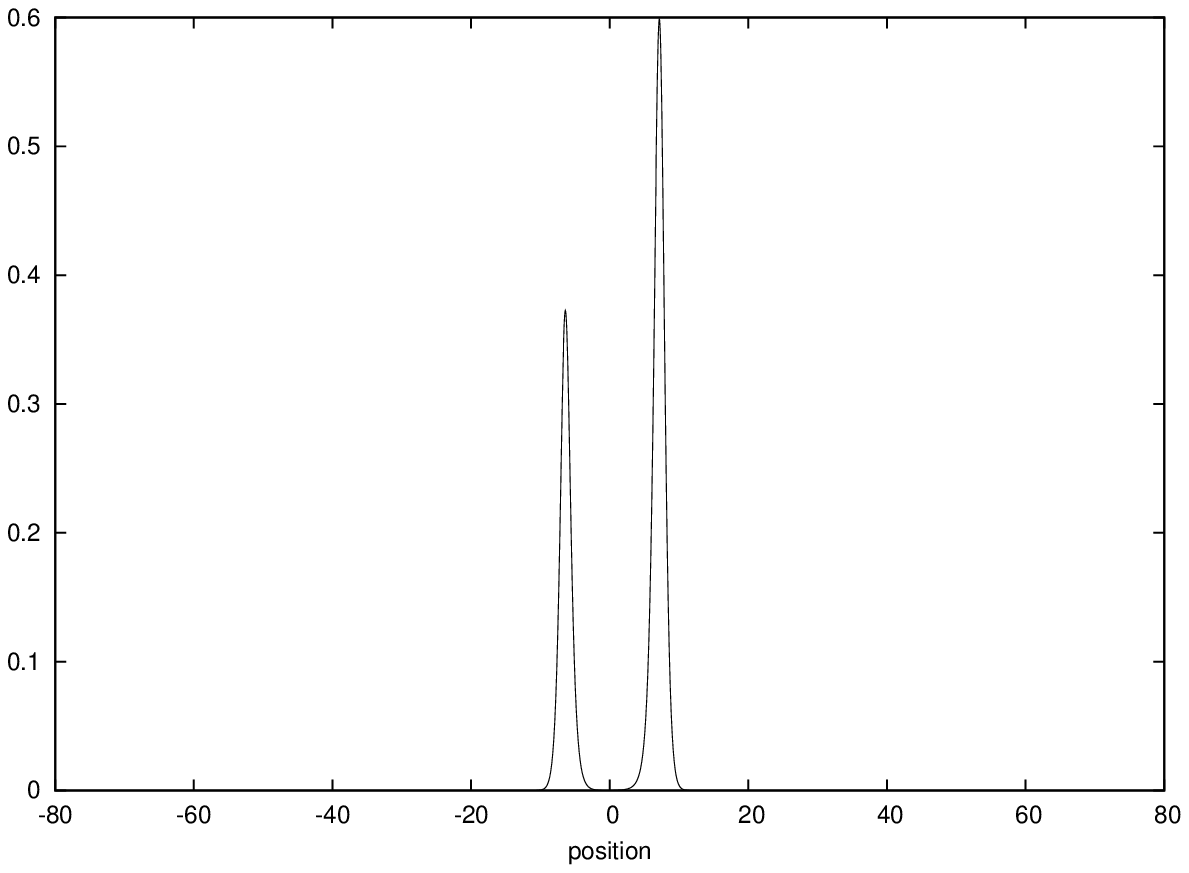}
\end{minipage}\vskip 0.4truecm 
\begin{minipage}[c]{0.51\textwidth}
\includegraphics[width=5.2cm]{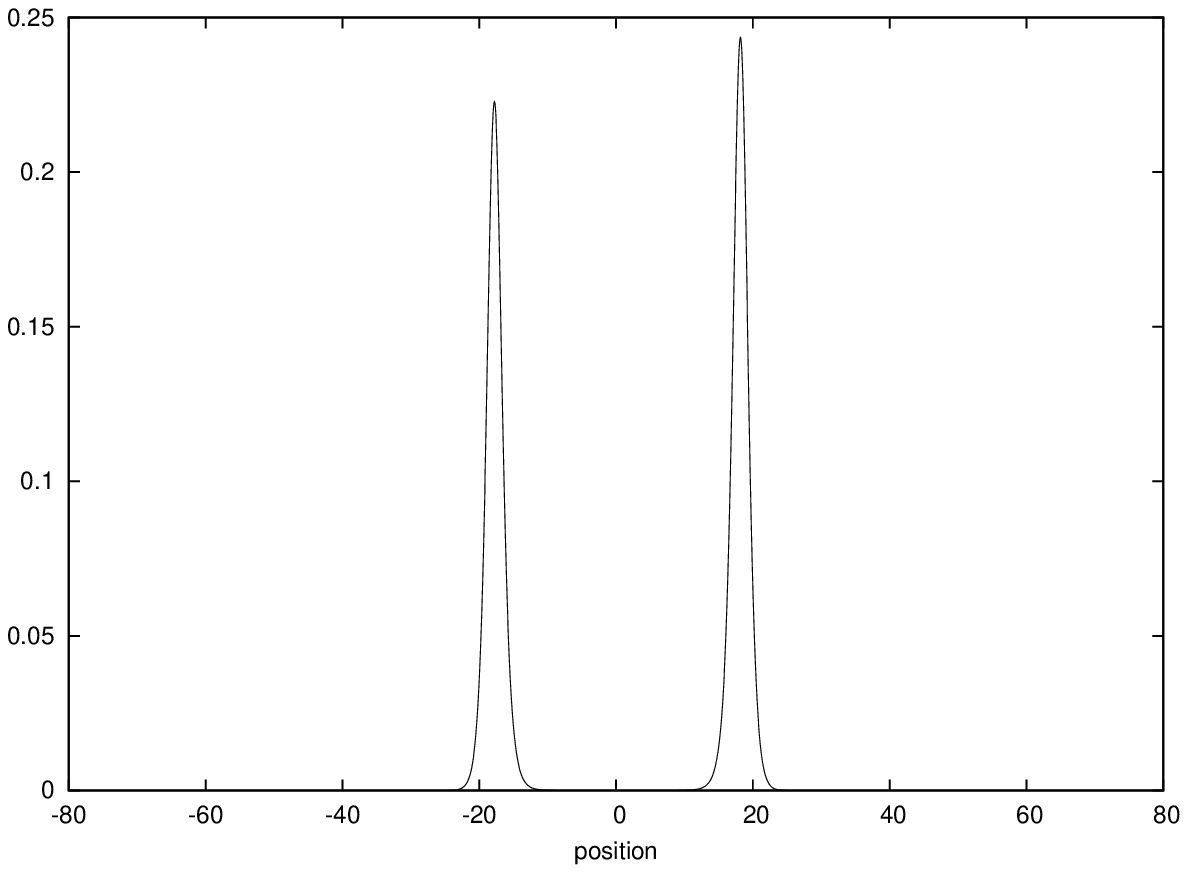}
\end{minipage}%
\begin{minipage}[c]{0.51\textwidth}
\includegraphics[width=5.2cm]{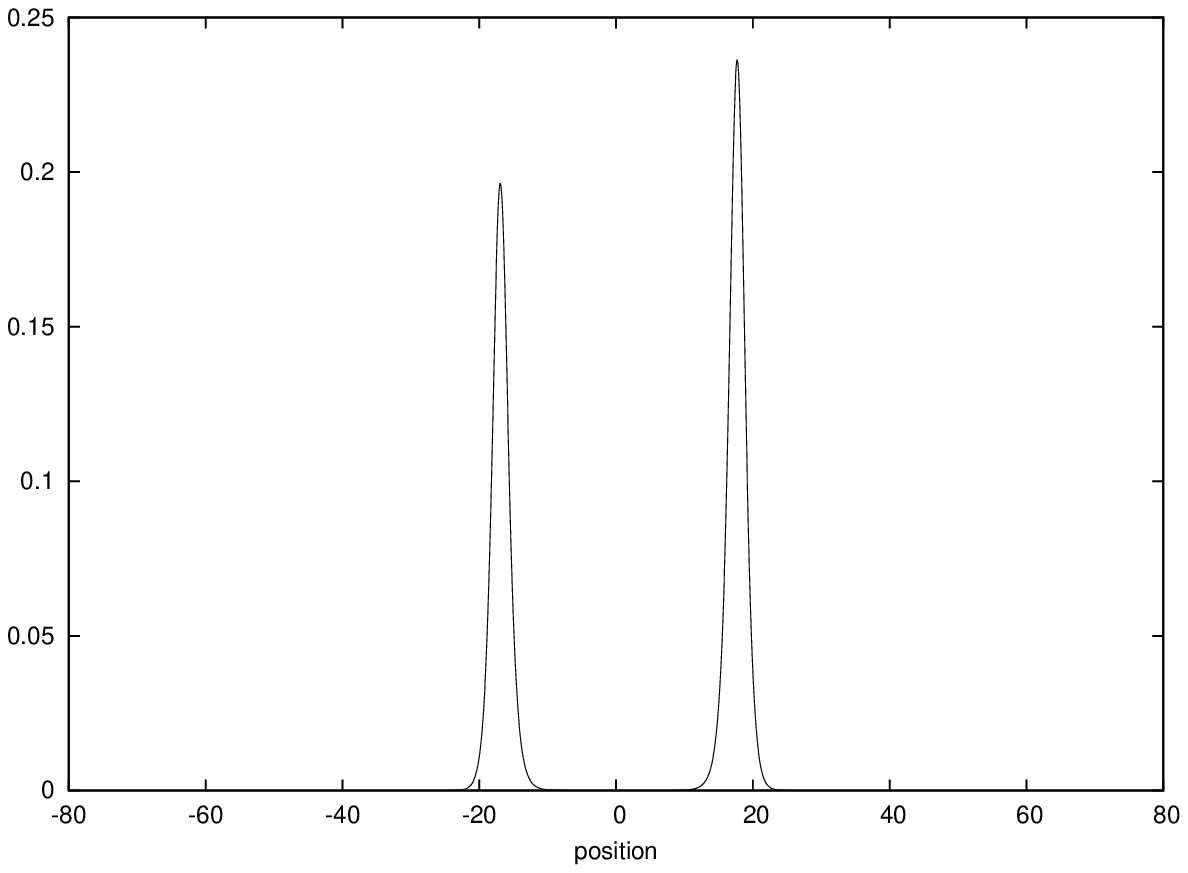}
\end{minipage}\vskip 0.4truecm
\begin{minipage}[c]{0.51\textwidth}
\includegraphics[width=5.2cm]{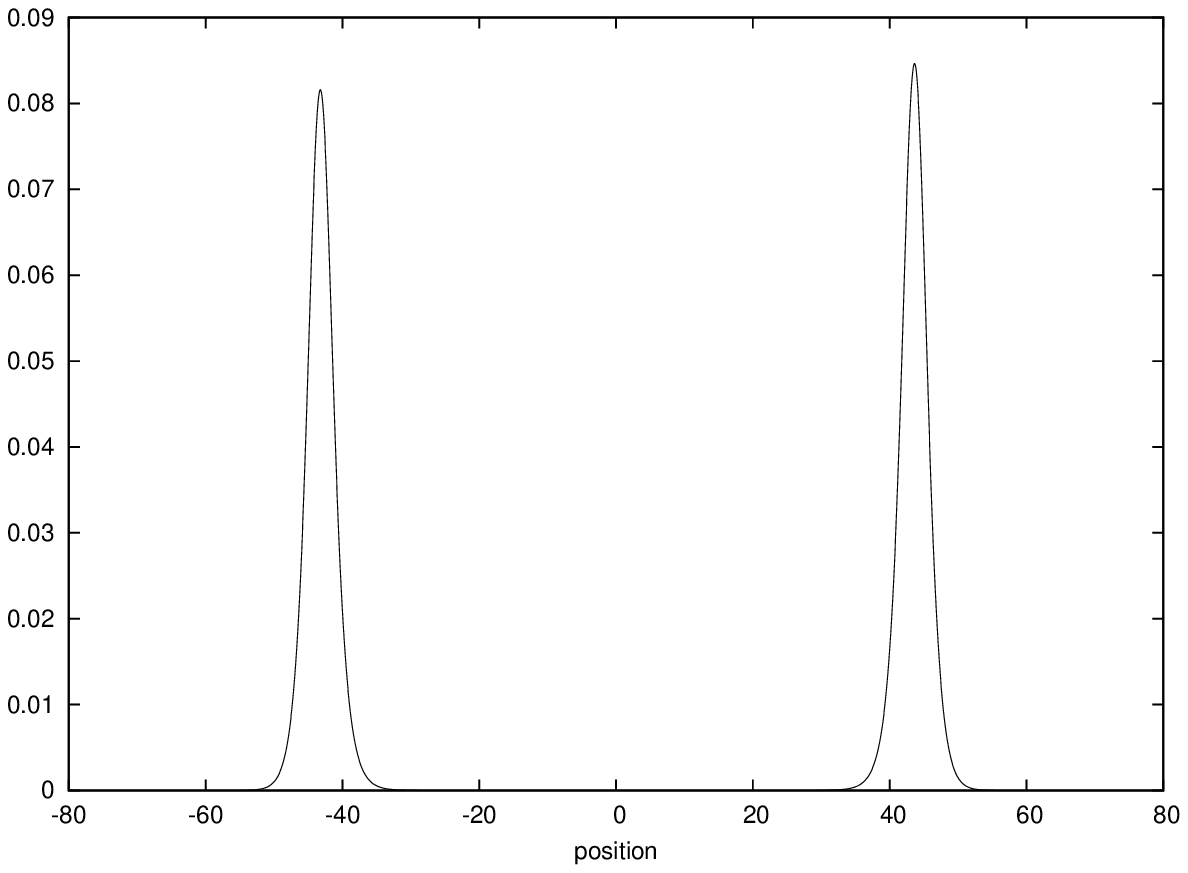}
\end{minipage}%
\begin{minipage}[c]{0.51\textwidth}
\includegraphics[width=5.2cm]{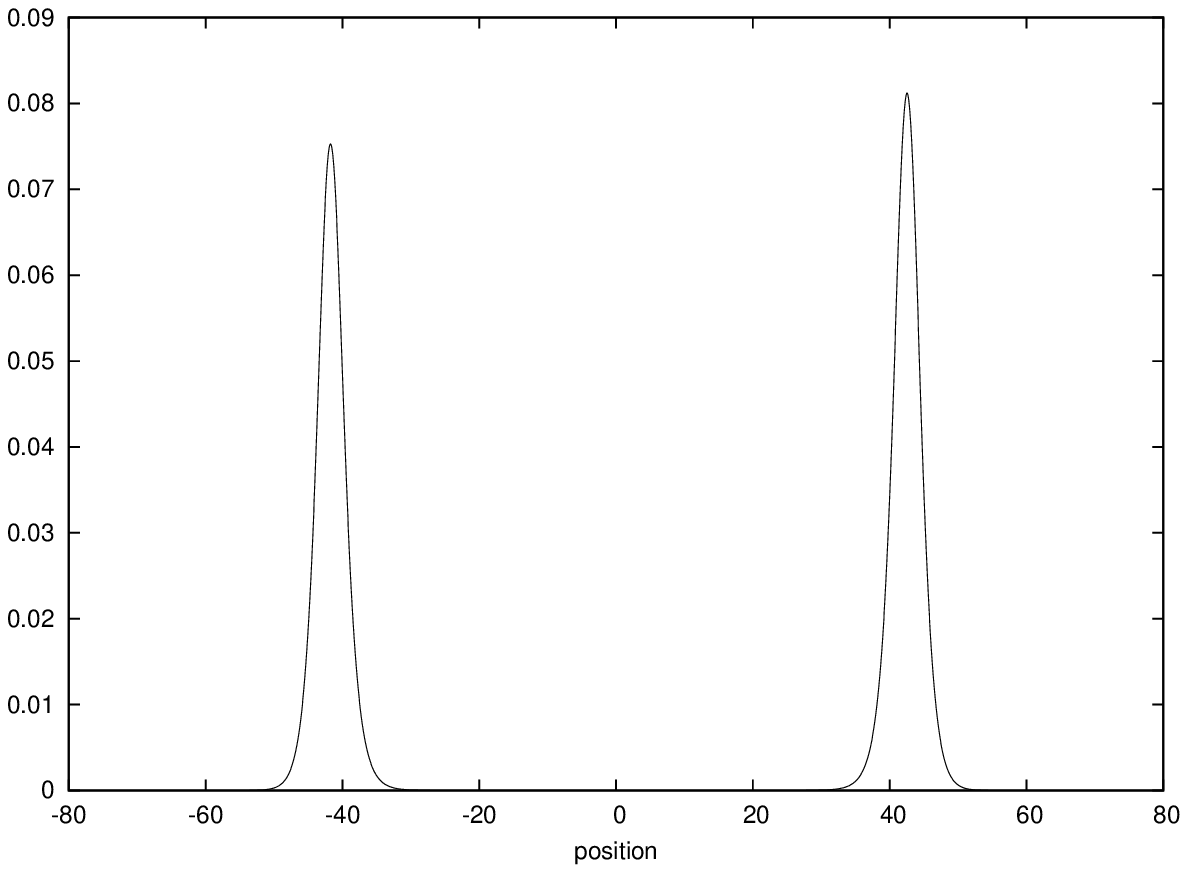}
\end{minipage}\vskip 0.4truecm 
\begin{minipage}[c]{0.51\textwidth}
\includegraphics[width=5.2cm]{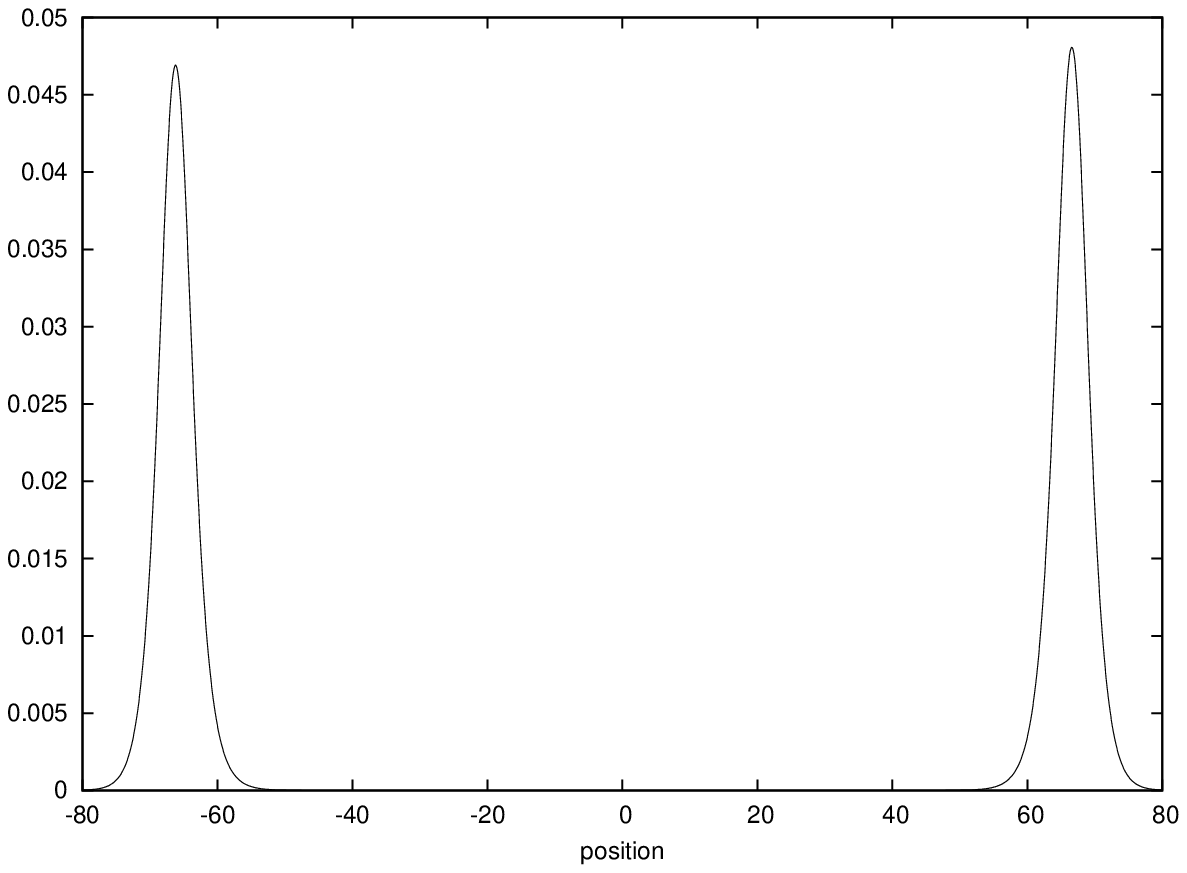}
\end{minipage}%
\begin{minipage}[c]{0.51\textwidth}
\includegraphics[width=5.2cm]{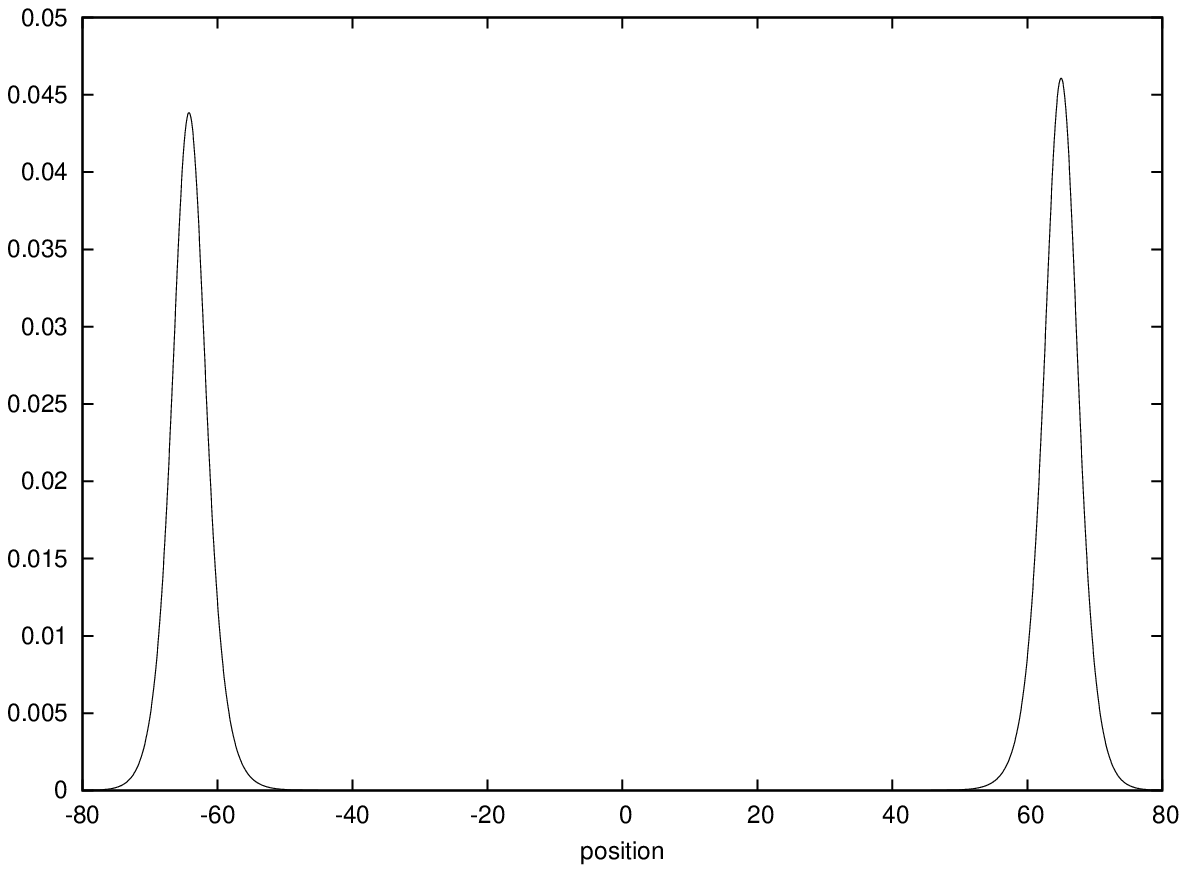}
\end{minipage} 
\caption{Spatial profile of the intensity $\vert \psi(z, t)\vert$ of the matter-wave bisoliton for $P_a=0.05$ (left column) and $P_a=0.1$ (right column). From the top to the bottom graphs times are choosen as follows, in arbitrary units: $t=0$, $1.5$, $2.5$, $3.5$, $4$} 
\label{figthree}
\end{figure*}

Consider an hypothetical Bose-Einstein condensation experiment aimed at creating an atom laser for a double-slit interferometer. Starting from an initial setup consisting of one single BEC with identical masses and an attractive mean-field interaction between the atoms, we switch on adiabatically an harmonic anti-trap potential to produce a smooth barrier hump (i.e. a non-flat potential barrier) that can cut the single BEC into two slices. This gives rise to two overlapping BECs, and due to the presence of a finite attractive mean-field interaction in the two condensates we generate twinned matter-wave pulses of equal amplitudes. Next, suppose that the created bisolitonic matter-wave structure is subjected to the gravity force, so that the two constituent pulses are accelerated one with respect to the other. Overall, the gravity potential will cause breaking of symmetry of the harmonic expulsive potential and from the viewpoint of the Gross-Pitaevskii Eq.~(\ref{a1}), we expect a direct incidence of this symmetry breaking on the matter-waves structures. Our goal in this section is to examine what the analytical solution to the Gross-Pitaevskii equation for the system just described, obtained in~(\ref{a4}) via the IST technique, predicts in terms of the influence of combined effects of the expulsive potential and gravity on the dynamics of the double condensate at different representative orders of magnitude of the atomic mass $m$. \\
First, we start with an analysis of the effects of the gravitational force on spatial profiles of the matter-wave bisoliton. Fig.~(\ref{figone}) displays the magnitude $\vert \psi(z, 0)\vert^2$ of the matter-wave intensity as given by~(\ref{a4}) at the initial time $t=0$, for a fixed value of the amplitude $m\omega_{\parallel}^2$ of the harmonic barrier (i.e. $m\omega_{\parallel}^2=0.5$) but different atomic weights. More precisely, from the first to the last graph values of $P_a$ (in arbitrary units) are $0$, $0.01$, $0.1$, $0.4$, $0.5$ and $0.8$ respectively. 
The first graph corresponds to a context where the gravity force is absent (i.e. $P_a=0$) and only the anti-trap potential is switched on. As one sees, the two pulses in the bisoliton structure~(\ref{a4}) form a pair of perfect twins, characterized by equal tails and widths and hence should carry same amount of energy. Now when we also switch on the gravitational field, the perfect twinning of pulses breaks down as it emerges through shapes of the matter-wave intensity in the five last graphs of fig.~(\ref{figone}). Particularly remarkable is the difference of amplitudes among the two pulses, which is more and more pronounced as the atomic weight increases. At sufficiently large values of $P_a$, one of the twin pulses seems to collapse reflecting a strong influnce of the gravity tending to confine all atoms within one of the two condensates. The last behaviour is samewhat comparable with the well-known phenomenon of macroscopic self trapping~\cite{shenoy}, but except that in the present context atoms are exchanged between the two overlapping condensates by means of classical motion while in the common macroscopic self-trapping phenomenon, the density of population imbalance is determined by quantum tunnelings via a Josephson-like phase dynamics~\cite{shenoy,liub,liuc}. It should equally be stressed that in real experiments, the account of gravity must favour non-symmetric positions of the two overlapping condensates with respect to a common axis coinciding with the axis of the gravity force. In such context the free fall process provides a relevant enhancing mehanism for atomic exchanges between the two condensates for heavy atoms that are expelled from an upper condensate to a down condensate by means of thermal ejections (i.e. thermal evaporation). \\
We also examined profiles of the moving twinned pair, still for different values of the atomic weight and keeping $m\omega_{\parallel}^2=0.5$. Since the integrals with respect to time in~(\ref{a5}) and~(\ref{a6}) are not tractabe analytically, we applied a numerical integration scheme assuming $t=0$ as the initial time. Figs.~(\ref{figtwo}) and~(\ref{figthree}) summarize variations of the matter-wave intensity in space at different times, for different atomic weights. 
On fig.~(\ref{figtwo}), the left column displays spatial profiles of the bisoliton intensity for $P_a=0$ at five different times namely $t=0$, $1.5$, $2.5$, $3.5$ and $4$ from the top graph respectively, while the right column are spatial profiles at same times but for an atomic weight $P_a=0.01$. Fig.~(\ref{figthree}) is a continuation of fig.~(\ref{figtwo}) where the left column corresponds to $P_a=0.05$ and the right column to $P_a=0.1$. According to curves of the two figures, in general the twin pulses composing the matter-wave bisoliton described by the analytical solution~(\ref{a4}) propagate in opposite directions. As the graphs for finite weights indicate, when $P_a$ increases the twin pulses aquire different amplitudes while still moving in opposite directions. At sufficiently long propagation time, their amplitudes both decrease and the bisoliton tends to recover its perfect twin structure. \\
An exact explanation of the apparent recovery of perfect twinning of the two pulses in the bisoliton after a long propagation time is not evident, as this behaviour emerges through numerical simulations of the full expression~(\ref{a4}) at finite times and positions. However, we suspect this behaviour to originate from a complex variation of the prefactor $\psi_0(z, t)$ when both time and space are simultaneously nonzero. In fact, the matter-wave amplitude is actually not constant but changes with space and time suggesting a non-topological feature for the bisoliton solution found in~(\ref{a4}).
\section{Conclusion}
In summary we developed a non-perturbative approach to the Gross-Pitaevskii equation in the presence of an harmonic non-symmetric expulsive potential, in order to gain a more consistent insight about the dynamics of BEC systems under combined effects of an expulsive potential and the gravitational field~\cite{coq,hope}. We found that the expulsive potential, which has shape of a smooth space-varying (i.e. harmonic) barrier, favors twinned matter-wave structures and thus acts like an atomic beam splitter. The gravity was found to break the symmetry of this harmonic barrier turning the perfect twin-pulse bisoliton into an asymmetric bisoliton. This asymmetry emerged through their distinct tails, widths and positions with respect to a common initial position along the gravity axis, and appeared to be more and more pronounced as the atomic weight was increased. However, while the spatial profile of bisoliton suggests a monotonic change in initial amplitudes of the two pulses, the space-time evolution instead reveals a complex behaviour involving possible recovery of the perfect pulse twinning after a sufficiently long propagation time. This behaviour has been infered to a non-topological (i.e. dynamical) feature of the bisoliton. \\
Dynamical properties of BECs under combination of the expulsive harmonic barrier and the gravity that emerge in the present study, reveal a quite rich physics some aspects of which have already be experimentally predicted~\cite{mewes,bloch,coq,hope,mori,shimizu}. The mathematical method adopted, namely the IST technique, clearly provides a framework where this rich physics can be captured far better than within perturbative approaches, or some variational treatments of the anti-trap and/or gravity.
\section*{Acknowledgments}
A. M. Dikand\'e thanks Yi Zhang (from the Zhejiang Normal University) for sharing their recent results on solutions to a close problem, following Hirota's Bilinear method.

\end{document}